\documentclass[JGR: Planets]{agujournal2019}
\usepackage[LGR,T1]{fontenc}
\usepackage{amsmath}
\usepackage{amssymb}
\usepackage{mhchem}
\usepackage{xfrac}
\usepackage{graphicx}
\usepackage{subfig}
\usepackage{cancel}
\usepackage{indentfirst}
\usepackage{caption}
\usepackage{gensymb}
\usepackage{multicol}
\usepackage{multirow}
\usepackage{pdflscape}
\usepackage{wasysym}
\usepackage{soul}
\usepackage[normalem]{ulem}
\journalname{JGR: Planets}

\begin{document}

%TC:ignore
\title{The Physics of Falling Raindrops in Diverse Planetary Atmospheres}

\authors{Kaitlyn Loftus\affil{1}, Robin D. Wordsworth\affil{1,2}}

\affiliation{1}{Department of Earth and Planetary Sciences, Harvard University, Cambridge, MA, US}
\affiliation{2}{School of Engineering and Applied Sciences, Harvard, University, Cambridge, MA, US}

\correspondingauthor{Kaitlyn Loftus}{kloftus@g.harvard.edu}

\begin{keypoints}
\item We present general methods to calculate raindrop shape, speed, and evaporation rate in diverse planetary atmospheres
\item  We define a dimensionless number that we show can closely capture the behavior of raindrop evaporation
\item Maximum stable raindrop size is relatively insensitive to  condensible species and atmospheric properties (including air density)
\end{keypoints}

\begin{abstract} 
The evolution of a single raindrop falling below a cloud is governed by fluid dynamics and thermodynamics fundamentally transferable to planetary atmospheres beyond modern Earth's. Here, we show how three properties that characterize falling raindrops---raindrop shape, terminal velocity, and evaporation rate---can be calculated as a function of raindrop size in any planetary atmosphere. We demonstrate that these simple, interrelated characteristics tightly bound the possible size range of raindrops in a given atmosphere, independently of poorly understood growth mechanisms. Starting from the equations governing raindrop falling and evaporation, we demonstrate that raindrop ability to vertically transport latent heat and condensible mass can be well captured by a new dimensionless number. Our results have implications for precipitation efficiency, convective storm dynamics, and rainfall rates, which are properties of interest for understanding planetary radiative balance and (in the case of terrestrial planets) rainfall-driven surface erosion. 
\end{abstract}

\section*{Plain Language Summary}
The behavior of clouds and precipitation on planets beyond Earth is poorly understood, but understanding clouds and precipitation is important for predicting planetary climates and interpreting records of past rainfall preserved on the surfaces of Earth, Mars, and Titan.  
One component of the clouds and precipitation system that can be easily understood is the behavior of individual raindrops. Here we show how to calculate three key properties that characterize raindrops: their shape, their falling speed, and the speed at which they evaporate. From these properties, we demonstrate that, across a wide range of planetary conditions, only raindrops in a relatively narrow size range can reach the surface from clouds. We are able to abstract a very simple expression to explain the behavior of falling raindrops from more complicated equations, which should facilitate improved representations of rainfall in complex climate models in the future.

%TC:endignore

\section{Introduction}\label{sec:intro}
Within a planetary condensible cycle, precipitation is the transport of the condensible species in a condensed phase (liquid or solid) through the atmosphere and, for terrestrial planets, to the surface. Extensive vertical displacement relative to the local air mass distinguishes precipitation from clouds. Because precipitating particles can fall far from the air mass where they form, they redistribute both heat and the condensible species within an atmosphere. Precipitation is a transient state, but though its effects are largely indirect, they have immense consequences for planetary climate.

The behavior of precipitation is essential to setting planetary radiative balance. Precipitation’s role in transporting condensible mass from the atmosphere to the surface (or the deep atmosphere on gaseous planets) exerts a strong influence on the relative humidity distribution \cite{Sun1993,Romps2014,Lutsko2018,Ming2018}, cloud lifetimes and occurrence rates \cite{Zhao2016,Seeley2019},
and condensible surface distributions \cite{Abe2011,Wordsworth2013a}. These properties, in turn, have direct radiative implications via the greenhouse effect and albedo changes \cite<e.g.,>[]{Pierrehumbert2007,Pierrehumbert2010,Shields2013,Yang2014,IPCC}. The role of precipitation in dictating radiative balance is especially important on dry planets \cite{Abe2011} and planets in or near a runaway greenhouse state \cite<e.g.,>[]{Pierrehumbert1995,Leconte2013}.

On Earth, global precipitation patterns play a critical role in determining local ecology and have significant societal impacts \cite{Margulis2017}. 
On terrestrial planets generally, the intensity, frequency, and spatial distribution of liquid precipitation are essential in governing surface erosion via runoff and physical weathering \cite<e.g.,>[]{Margulis2017} as well as chemical weathering fundamental to the carbon-silicate cycle \cite{Walker1981,Macdonald2019,Graham2020}. 
Finally, interpreting solar system geological records shaped by fluvial erosion---e.g., ancient Mars' large-scale valley networks and crater modifications \cite<e.g.,>[]{Craddock2002}, modern Titan's lakes and rivers \cite<e.g.,>[]{Lorenz2008}, and Archean Earth's fossilized raindrops \cite{Som2012,Kavanagh2015}---requires an understanding of changes in precipitation events as planetary conditions vary. 

Despite the importance of precipitation, little progress has been made on how precipitation physically behaves in different planetary environments \cite<e.g.,>[]{Vallis2020}. 
Previous studies have tended to view precipitation behavior primarily as a function of cloud formation and evolution. Cloud physics is complicated by extreme nonlinearities, gaps in theoretical understanding of fundamental processes bridged only by empiricisms, and dependencies on spatial and temporal scales that span many orders of magnitude. In turn, planetary clouds are studied via a hierarchy of models with varying tradeoffs among complexity, robustness of included processes, and ease of interpretation \cite<e.g.,>[]{Rossow1978,Carlson1988,Ackerman2001,Lee2016,Powell2018,Gao2018}.  
Still, climate models of terrestrial planets commonly represent clouds and precipitation using modern-Earth-tuned parameterizations with essentially ad hoc parameter sweeps \cite<e.g.,>[]{Wordsworth2013a,Urata2013,Yang2014,Komacek2019}.

Given the transition from cloud to precipitation is complex and poorly modeled even on the well-observed modern Earth \cite<e.g.,>[]{Rogers1996,Pruppacher2010,Flato2014}, any insights on precipitation behavior become almost impossible to extract when precipitation is considered only as an afterthought of cloud behavior. An alternative approach is to consider the behavior of individual precipitating particles independently of their formation conditions. This strategy is much more tractable in different planetary environments because individual precipitating particles are governed by thermodynamics and fluid dynamics that are both relatively well understood and fundamentally transferable to planetary regimes beyond modern Earth's. \citeA{Lorenz1993} took this approach and used key properties of individual methane-nitrogen raindrops on Titan to highlight fundamental differences between rainfall on Titan and Earth and hypothesize consequences for storm intensities. 

Previous planetary science studies have also attempted to use the simplicity of raindrop physics to place constraints on paleo-air pressures on Archean Earth \cite{Som2012,Kavanagh2015} and early Mars \cite{Craddock2017,Palumbo2020} via maximum raindrop sizes before breakup. This use of raindrop physics hints at the possible productivity of this approach. However, even maximum raindrop size has not been considered systematically in general planetary atmospheres before, and the conclusions reached by these studies have in some cases been inconsistent. Specifically, a recent study by \citeA{Craddock2017} reached opposite conclusions versus other studies \cite<e.g.,>[]{Komabayasi1964,Lorenz1993,Clift2005,Pruppacher2010,Som2012,Kavanagh2015,Palumbo2020} on the dependence of maximum stable raindrop size on planetary air density, leading to entirely different conclusions about paleo-air densities. 

In this paper, we build on \citeA{Lorenz1993} to establish a comprehensive and generalized picture of the ``life and death'' of a single raindrop over a wide range of planetary conditions. Like \citeA{Lorenz1993}, we neglect the ``birth'' of the raindrop (i.e., the growth of a cloud particle of negligible vertical velocity to a precipitating particle below a cloud, essentially done growing). Our approach here is distinct from previous work as we present fully generalized methods, tie each component of our methodology back to fundamental physics, and focus on how the well-understood behavior of an individual raindrop can provide insight into the rest of the condensible cycle in different planetary environments. This work lays a foundation for building physically driven microphysics parameterizations for generalized mesoscale models and global circulation models (GCMs).

We limit our focus to liquid precipitating particles (``raindrops'') because they have a unique shape for a given mass of condensible. The shape degeneracy of solid precipitating particles is a major challenge \cite<e.g.,>[chapter 2.2]{Pruppacher2010} that we do not treat here.
However, once the non-uniqueness of solid precipitating particle shapes is addressed, our methodology is applicable to solid particles as well. Water is the most familiar condensible species, but all methodology we present here is generalized for any liquid condensible, e.g., methane-nitrogen raindrops on Titan or iron raindrops on WASP-76b \cite{Ehrenreich2020}. Except when we assume the existence of a planetary surface, our methodology is also general to both terrestrial and gaseous planets.

In section \ref{sec:methods1} we present methods to calculate falling raindrop shape, terminal velocity, and evaporation rate in a generic atmosphere. We show how these characteristics can place upper and lower bounds on raindrop size in section \ref{sec:methods2}. Section \ref{sec:results} uses the methodology developed to probe raindrop characteristics and size bounds in different atmospheres.
In section \ref{sec:diss} we discuss the implications of our results for different planetary atmospheres and microphysics parameterizations as well as possible extensions of this work to solid precipitating particles. 
We summarize our results in section \ref{sec:con}.

\section{Raindrop Characteristics}\label{sec:methods1}
In isolation, a precipitating particle does two things:  (1) it falls and (2) it evaporates. To calculate the rate at which a particle falls and evaporates requires knowledge of the relationship between particle mass and shape. Unlike solid precipitating particles, whose forms vary widely, raindrops have equilibrium shapes that can be uniquely calculated for a given mass of liquid condensible, air density, and surface gravity. A unique shape allows us, in a known external environment, to describe a raindrop with only a single size variable. Here, we use equivalent radius $r_\text{eq}$, which is the radius a raindrop of mass $m$ would have if it were spherical, i.e.,
\begin{linenomath*}
\begin{equation}
    m = \frac{4}{3}\pi\rho_{\text{c},\ell}r_\text{eq}^3
\end{equation}
\end{linenomath*}
where $\rho_{\text{c},\ell}$ is the density of the liquid condensible.

\subsection{Raindrop Shape}\label{subsec:r_shape}
Falling raindrops adopt a range of shapes depending on their size---though never the teardrop shape inscribed in the public imagination \cite{Blanchard2004}. As raindrops grow in mass, they evolve from spheres to oblate spheroids to shapes resembling the top of a hamburger bun \cite<e.g.,>[]{Pruppacher1971,Beard1987}. (An oblate spheroid is generated from an ellipse rotated about its minor axis.) Spheres are merely a specific subset of oblate spheroid, and the more complex shapes have virtually indistinguishable dynamical properties from oblate spheroids \cite{Green1975,Beard1987,Szakall2010}; therefore, we simplify our shape calculations by prescribing that raindrops take the shape of an oblate spheroid with semi-major axis $a$ (oriented perpendicular to the fall direction) and semi-minor axis $b$ (oriented parallel to the fall direction). Given this assumption, we can describe raindrop shape with only an axis ratio $b/a$ and make use of many existing analytic expressions. Geometric properties of oblate spheroids used in this paper are given in \ref{app:oblate}. 

In equilibrium, the deviation of a raindrop surface from a minimum-energy-state sphere can be calculated considering the first law of thermodynamics:  the change in energy from the surface's increased surface tension must be balanced by the work done to expand the surface's enclosed volume into a region of different pressure \cite<e.g.,>[chapter 10.3.2]{Pruppacher2010}. We follow \citeA{Green1975} in accounting for the key pressures at raindrop equator and assuming an oblate spheroid shape (see \ref{app:shape} for more detail), which gives 
\begin{linenomath*}
\begin{equation}\label{eq:shape}
    r_\text{eq} = \sqrt{\frac{\sigma_{\text{c-air}}}{g (\rho_{\text{c,}\ell}- \rho_\text{air})}}\left (\frac{b}{a} \right)^{-\frac{1}{6}} \sqrt{\left (\frac{b}{a} \right)^{-2} -2 \left(\frac{b}{a} \right)^{-\frac{1}{3}} + 1}.
\end{equation}
\end{linenomath*}
Here $\sigma_{\text{c-air}}$ is the surface tension between the liquid condensible and air, $g$ is the local gravitational acceleration, and $\rho_\text{air}$ is the local air density. 
For a given $r_\text{eq}$, equation \eqref{eq:shape} can be solved numerically for $b/a$ to characterize the raindrop's shape. 

\subsection{Raindrop Velocity}\label{subsec:v}
The two key forces acting on a falling raindrop in a planetary atmosphere are the gravitational force $F_\text{g}$ and the aerodynamic drag force $F_\text{drag}$.
The effective gravitational force on a raindrop is 
\begin{linenomath*}
\begin{equation}\label{eq:f_g}
    F_\text{g} = \frac{4}{3}\pi r_\text{eq}^3(\rho_{\text{c,}\ell}-\rho_\text{air})g
\end{equation}
\end{linenomath*}
after accounting for the raindrop's buoyancy within the air fluid.
Drag force on a raindrop is 
\begin{linenomath*}
\begin{equation}\label{eq:f_drag}
    F_\text{drag} = \frac{1}{2}C_\text{D}A\rho_\text{air}v^2
\end{equation}
\end{linenomath*}
where $C_\text{D}$ is the drag coefficient of the raindrop, $A$ is cross sectional area of the raindrop, $\rho_\text{air}$ is the local air density, and $v$ is the raindrop's fall speed relative to air. 
 Raindrop cross sectional area is a function of raindrop shape and size. 
 $C_\text{D}$ is a function of raindrop shape and flow regime. The latter can be characterized by the dimensionless Reynolds number Re 
\begin{linenomath*}
\begin{equation}
    \text{Re} \equiv \frac{v\ell\rho_\text{air}}{\eta_\text{air}}
\end{equation}
\end{linenomath*}
where $\ell$ is a characteristic raindrop length scale that we take to be $2r_\text{eq}$ and $\eta_\text{air}$ is the dynamic viscosity of local air. Viscosity varies with both air composition and temperature;
we calculate $\eta_\text{air}$ in a generic air mixture following \citeA{Reid1977} chapter 9. 

Calculating $C_\text{D}$ theoretically for a general-shaped object in a general flow is not tractable \cite<e.g.,>[]{Stringham1969}, and thus $C_\text{D}$ is typically evaluated via experimentally based parameterizations. As long as the flow regime is correctly captured by a scale analysis, defaulting to experimentally based parameterizations for generic atmospheric conditions in and of itself is not problematic---though limitations in the coverage of the parameter space used to fit expressions must be considered. 

In this work, we use the drag parameterization
\begin{linenomath*}
\begin{equation}\label{eq:C_D}
    C_D = \left(\frac{24}{\text{Re}}\left(1 + 0.15 \text{Re}^{0.\overline{687}}\right) + 0.42\left(1 + 4.25\times10^{4}\text{Re}^{-1.16}\right)^{-1}\right)C_\text{shape}
\end{equation}
\end{linenomath*}
\cite{Clift1970,Loth2008}.
$C_\text{shape}$ is a correction term to account for raindrop deviations from spherical shape fit by \citeA{Loth2008} for $b/a \le 1$ across a variety of falling object shapes:
\begin{linenomath*}
\begin{equation}\label{eq:C_shape}
    C_\text{shape} = 1 + 1.5\left(f_\text{SA}-1\right)^{0.5} + 6.7(f_\text{SA}-1)
\end{equation}
\end{linenomath*}
where $f_\text{SA}$ is the ratio of the surface area of the oblate spheroid raindrop to the surface area of a sphere of radius $r_\text{eq}$. The formulation of $C_D$ in equation \eqref{eq:C_D} is primarily based on the parameterization of drag for a sphere with  $\text{Re}<3.5\times10^5$ by \citeA{Clift1970}. $C_\text{shape}$ is most accurate when significant raindrop deviation from a sphere occurs within the Newtonian flow regime of Re between about 750 to 3.5$\times10^5$ \cite{Clift2005,Loth2008}. 

Compared to the other $C_D$ parameterizations we considered \cite{Salman1988,Lorenz1993,Ganser1993,Holzer2008}, we found equation \eqref{eq:C_D} with \eqref{eq:C_shape} best reproduced the velocity dependence on raindrop radius for Earth values with experimental validation over wide ranges of Re and oblate spheroid axis ratios.
We note that we neglect corrections to $C_D$ for non-continuum regime effects (commonly referred to as the Cunningham or slip-flow correction factor) as we are not concerned with the behavior of very small ($\lesssim1$ \textmu m) particles \cite<e.g.,>[chapter 9.2]{Seinfeld2006}. (Such corrections only become important when particle size becomes comparable to the mean free path of local air molecules.)

Terminal velocity $v_\text{T}$ occurs when the raindrop is no longer accelerating, and the gravitational force $F_\text{g}$ is balanced by the aerodynamic drag force $F_\text{drag}$. Under modern Earth atmospheric conditions, the timescale for raindrops of a fixed size to reach terminal velocity is very small compared to their lifetimes \cite<>[chapter 10.3.5]{Pruppacher2010}. 
% page 327-328 of PK 
To test the generality of this rapid raindrop acceleration assumption, we numerically integrated water raindrop motion accounting for variable acceleration across a large range of plausible planetary conditions (air composition, surface gravity, air pressure, air temperature). 

We find that the Earth-based empiricism is generally true: even the largest possible stable raindrops (described in section \ref{subsec:r_max}) with the longest acceleration timescales reach 99\% of their terminal velocity after starting from rest within the first 1\% of their total fall distance and 5\% of their total fall time (Figure S1). Further, when considering the effect of raindrop size changes due to evaporation on reaching terminal velocity, we find that the differences in raindrop fall speed between self-consistent treatment of raindrop acceleration and assuming terminal velocity is instantly reached are, at maximum, on the order of 10\% and typically much smaller (Figure S2).
Thus, for simplicity, we henceforth make the standard assumption that raindrop falling speed relative to air $v$ is the raindrop's terminal velocity, which can be uniquely determined for a given raindrop size and shape. 

Equating $F_\text{drag}$ and $F_g$ and substituting the appropriate oblate spheroid geometry yields a terminal velocity of
\begin{linenomath*}
\begin{equation}\label{eq:v}
    v_\text{T} = -\sqrt{\frac{8}{3}\frac{(\rho_{{\text{c,}\ell}}-\rho_\text{air})}{\rho_\text{air}}\frac{g}{C_D}\left(\frac{b}{a}\right)^{\frac{2}{3}}r_\text{eq}}.
\end{equation}
\end{linenomath*}
$\rho_\text{air}$ and $g$ are known from planetary atmospheric properties. $b/a$ is uniquely determined from $r_\text{eq}$ with equation \eqref{eq:shape}. The nonlinear dependence of $C_D$ on $v$ (through Re) requires we solve equation \eqref{eq:v} numerically.

A raindrop's vertical speed relative to a planet's surface---d$z$/d$t$, its change in altitude $z$ per unit time $t$---is the sum of the its velocity relative to air, here assumed to be $v_\text{T}$, and the vertical velocity of the raindrop's local air $w$:
\begin{equation}\label{eq:dzdt_raindrop}
    \frac{\text{d}z}{\text{d}t} = v_\text{T} + w.
\end{equation}
We are focused on raindrop physics here and so treat $w$ as a free parameter in the analysis that follows.

\subsection{Evaporation Rate}\label{subsec:evaprate}
Raindrop evaporation occurs when the atmosphere surrounding the drop is sub-saturated in condensible gas. The preferred phase of the condensible molecules at the drop's surface becomes gas rather than liquid. As the condensible is transferred from the liquid phase in the raindrop to the gas phase in the air, the air closest to the raindrop surface deviates in temperature and relative humidity from the local atmospheric state; the relative humidity adjacent to the drop surface increases while the temperature drops due to the latent heat required for the liquid-to-gas phase transition. Both these effects serve to lower the thermodynamic impetus to evaporate. Thus, in addition to the environmental level of sub-saturation, the rate at which evaporation occurs is dictated by the rate at which heat and condensible gas can be transported away from the raindrop surface.

Quantitatively, the change in raindrop equivalent radius
with time $t$ can then be formulated from geometry and appropriate boundary conditions as
\begin{linenomath*}
\begin{equation}\label{eq:drdt}
    \frac{\text{d}r_\text{eq}}{\text{d}t} = \frac{f_\text{V,mol} D_\text{c-air}\mu_\text{c}}{r_\text{eq}\rho_{\text{c,}\ell}R}\left(\text{RH}\frac{p_\text{c,sat}(T_\text{air})}{T_\text{air}} - \frac{p_\text{c,sat}(T_\text{drop})}{T_\text{drop}}\right)
\end{equation}
\end{linenomath*}
\cite<see>[chapter 7 for a derivation]{Rogers1996}. $\text{RH}$ is the relative humidity of the local air; $R$ is the ideal gas constant; $\mu_\text{c}$ is the molar mass of the condensible in its gas phase; $\rho_{\text{c,}\ell}$ is the density of the liquid condensible; $D_\text{c-air}$ is the diffusion coefficient for the condensible gas in air; $p_\text{c,sat}$ is condensible gas saturation pressure; $T_\text{air}$ is the local air temperature far from the drop's surface; $T_\text{drop}$ is the temperature at the raindrop's surface; and $f_\text{V,mol}$ is a ventilation factor that accounts for how much raindrop motion enhances condensible molecule transport relative to a stagnant drop.

Conservation of heat at the raindrop surface yields the following differential equation governing $T_\text{drop}$:
\begin{linenomath*}
\begin{equation}\label{eq:dTdt}
     \frac{\text{d}T_\text{drop}}{\text{d}t} = \frac{3}{r_\text{eq} c_{p\text{,c,}\ell}} \left(L_\text{c} \frac{\text{d}r_\text{eq}}{\text{d}t} - \frac{f_\text{V,heat}K_\text{air}}{\rho_{\text{c,}\ell} r_\text{eq}}(T_\text{drop}-T_\text{air})\right)
\end{equation}
\end{linenomath*}
\cite<>[chapter 7]{Rogers1996}. $c_{p\text{,c,}\ell}$ is the specific heat at constant pressure of the liquid condensible; $L_\text{c}$ is the condensible's latent heat of vaporization at $T_\text{drop}$; $K_\text{air}$ is thermal conductivity of air; $f_\text{V,heat}$ is a ventilation factor that accounts for how much raindrop motion enhances heat transport relative to a stagnant drop. Without any further simplifications, equations \eqref{eq:drdt} and \eqref{eq:dTdt} must be solved together numerically from initial conditions given their mutual dependencies. (We will discuss simplifications to calculating $T_\text{drop}$ in detail in section \ref{subsec:nond_num}.)

Relative humidity RH (also known as saturation) is defined as the ratio of the local condensible gas partial pressure $p_\text{c}$ to $p_\text{c,sat}$ at $T_\text{air}$, i.e,  $\text{RH}\equiv p_\text{c}/p_\text{c,sat}(T_\text{air})$.
$D_\text{c-air}$ is a function of temperature, pressure, and air composition that we calculate following \citeA{Reid1977} chapter 11 and \citeA{Fairbanks1950}. $K_\text{air}$ is a function of temperature as well as air composition that we calculate following the Eucken method \cite<>[chapter 10]{Reid1977}.
Note that the formulation of raindrop temperature in equation \eqref{eq:dTdt} only considers heat transport via conduction. For high temperature condensibles, heat transport via radiation will also need to be considered.

The ventilation factors arise from fluid dynamical effects not analytically calculable, so, as for $C_D$, we must evaluate $f_\text{V,mol}$ and $f_\text{V,heat}$ from parameterizations based on experiments.
Here, we choose to use the $f_\text{V,mol}$ parameterization of \citeA{Beard1971} and \citeA{Pruppacher1979}:
\begin{linenomath*}
\begin{equation}\label{eq:f_vmol}
f_\text{V,mol} = 
\begin{cases}
   1 + 0.108\left(\text{Re}^{0.5}\text{Sc}^{0.\bar{3}}\right)^2, & \text{Re}^{0.5}\text{Sc}^{0.\bar{3}} < 1.4, \\
   0.78 + 0.308\left(\text{Re}^{0.5}\text{Sc}^{0.\bar{3}}\right), & \text{Re}^{0.5}\text{Sc}^{0.\bar{3}} \ge 1.4
\end{cases}
\end{equation}
\end{linenomath*}
where Sc is the dimensionless Schmidt number defined as 
\begin{linenomath*}
\begin{equation}
    \text{Sc} \equiv \frac{\eta_\text{air}}{D_\text{c-air}\rho_\text{air}}.
\end{equation}
\end{linenomath*}
This parameterization is only experimentally validated for $\text{Re} < 2600$ but is hypothesized to be valid for spheres with $\text{Re} < 8\times10^{4}$ based on theory \cite{Pruppacher1979}. Following \citeA<>[chapter 13.2.3]{Pruppacher2010}, we calculate $f_\text{V,heat}$ from equation \eqref{eq:f_vmol} for $f_\text{V,mol}$ with Sc replaced by the mathematically analogous dimensionless Prandtl number Pr
\begin{linenomath*}
\begin{equation}
    \text{Pr} \equiv \frac{\eta_\text{air}c_{p\text{,air}}}{K_\text{air}}
\end{equation}
\end{linenomath*}
where $c_{p\text{,air}}$ is the specific heat at constant pressure for air. We neglect the effects of turbulence, which can act to increase ventilation.

Raindrop shape impacts multiple aspects of ventilation, but in raindrop experimental data considered by \citeA{Pruppacher1979}, these shape effects cancel each other such that $f_V$ is independent of raindrop shape deformations at larger Re. (Note that this shape independence is only true for liquid raindrops, not solid condensibles \cite<e.g.,>[chapter 13.3.2]{Pruppacher2010}.)
%pg 440, 450 (ice) in PK 

The above expressions for $f_\text{V,mol}$ and $f_\text{V,heat}$ are not definitive and could be improved by further experiments over a broader parameter space.
The ventilation factors increase d$r_\text{eq}$/d$t$ by roughly an order of magnitude at large raindrop sizes, so they need to be accounted for despite the somewhat limited coverage of these parameterizations; as we will show in section \ref{sec:results}, the qualitative impact of these uncertainties in ventilation for $r_\text{eq}(t)$ is often limited because large raindrops evaporate mass very gradually.

In our formulation of d$r_\text{eq}$/d$t$ in equation \eqref{eq:drdt}, we have made a number of simplifying assumptions that are valid here because we are concerned only with evaporation. (d$r_\text{eq}$/d$t$ can describe both evaporation and condensation depending on whether RH is less than or greater than 1.) Our focus on evaporation, rather than condensation, means we are not concerned with the behavior of very small drops, which, as we will show in Section \ref{sec:results}, always evaporate rapidly compared to larger drops.
We neglect corrections to relative humidity at the raindrop's surface due to surface tension and condensation nuclei solute effects---commonly known as Kelvin and Raoult effects, respectively \cite<e.g.,>[chapter 6]{Lohmann2016}---that are important only at very small radii ($r\lesssim1$ \textmu m). Note that in atmospheres where a gas component is soluble in the liquid condensible \cite<e.g., \ce{N2} in \ce{CH4} on Titan;>[]{Thompson1992}, solute corrections to RH cannot be neglected. (See, for example, \citeA{Graves2008} for how to extend what is presented here to such cases.) Consistent with our treatment of drag, we also neglect corrections to condensible vapor diffusivity and air thermal conductivity for non-continuum regime effects very close to the drop's surface, which is effectively equivalent to assuming that the mean free paths of air and condensible gas molecules are small compared to the size of the raindrop \cite<e.g., see>[chapter 8.2.2 for further discussion]{Lamb2011}. 

Beyond corrections only necessary for small drops, we also neglect corrections to the boundary conditions used to solve for d$r_\text{eq}$/d$t$ due to raindrop deformation from a sphere \cite<>[chapter 8.3]{Lamb2011}. These shape corrections considerably complicate manipulating d$r_\text{eq}$/d$t$ and cause variations in d$r_\text{eq}$/d$t$ of less than 5\%.

\section{Raindrop Size Constraints}\label{sec:methods2}

\subsection{Evolution of Raindrop Size with Height below a Cloud}\label{subsec:r_z}
From the raindrop characteristics we outlined in section \ref{sec:methods1}, we can calculate the change in raindrop equivalent radius with altitude $z$ as 
\begin{linenomath*}
\begin{equation}\label{eq:drdz}
    \frac{\text{d}r_\text{eq}}{\text{d}z} = \frac{\text{d}r_\text{eq}}{\text{d}t} \left(\frac{\text{d}z}{\text{d}t} \right)^{-1}
\end{equation}
\end{linenomath*}
where d$r_\text{eq}$/d$t$ is given by equation \eqref{eq:drdt} and d$z$/d$t$ by equation \eqref{eq:dzdt_raindrop}. Solving equation \eqref{eq:drdt} also requires coupled evaluation of d$T_\text{drop}$/d$t$ from equation \eqref{eq:dTdt}. Note that this formulation neglects the time to accelerate to a new terminal velocity as raindrop size changes (which is rapid compared to the timescale on which $r_\text{eq}$ evolves, as discussed in section \ref{subsec:v}).

To evaluate equation \eqref{eq:drdz}, we first must describe the atmospheric state variables that affect parameters required for calculating evaporation rate and terminal velocity---$p$, $T$, RH---as functions of $z$. In this work, we prescribe planetary conditions of $p$, $T$, and RH, at a single $z$---either at the surface or at the cloud base depending on the calculation of interest. We also require planetary inputs of $g$ and dry air composition, which are assumed to be fixed. 

We then follow the standard assumptions for a 1D atmosphere in radiative-convective equilibrium below saturated regions to relate atmospheric properties \cite<e.g.,>[]{Pierrehumbert2010,Romps2017}: our pressure-temperature profile follows a dry adiabat, $z$ is related to $p$ (and thus RH and $T$) assuming hydrostatic equilibrium, and RH is prescribed assuming the condensible gas is well-mixed (i.e., a constant molar concentration). Note that assuming constant $T$ and RH from average values below the cloud base does not lead to significantly different $\text{d}r_\text{eq}/\text{d}z$ values, but such a simple profile does not allow for an internally consistent calculation of cloud base height.  

We define cloud base as the ``lifting condensation level'' (LCL), the height at which a condensible gas reaches saturation in a parcel of air rising adiabatically:  $z$ such that $p_c(z_\text{LCL}) = p_{c,\text{sat}}(T(z_\text{LCL}))$. 
The LCL errs in predicting cloud base when limited cloud condensation nuclei require supersaturation to initiate cloud particle formation. However, as we are concerned with where the raindrop starts evaporating (which requires RH < 100\%), this caveat does not concern us. 

Equation \eqref{eq:drdz} is stiff, so we integrate it using an implicit Runge-Kutta method of order 5. We define the raindrop's initial $r_\text{eq}$ at cloud base as $r_0$. We calculate $r_\text{eq}$($z$) from the cloud base ($z = z_\text{LCL}$) to a desired $z$ or until the raindrop fully evaporates. Here we define a ``fully evaporated'' raindrop as a drop of equivalent radius less than a threshold drop size $\Delta r$.

We set $\Delta r$ = 1 \textmu m; our results are not sensitive to this choice of $\Delta r$ as long as $\Delta r \ll r_0$. $\Delta r$ must be non-zero for numerical stability, but $\Delta r>0$ also reflects the physical reality that the cloud drops that form raindrops very strongly thermodynamically favor condensing onto preexisting nuclei rather than forming homogeneously.

\subsection{Minimum Cloud-edge Raindrop Size to Reach a Given Height}
To understand the potential of a raindrop to transport condensible mass and heat within an atmosphere, we calculate the cloud-edge (where RH transitions to less than 1 and evaporation begins) size threshold where a raindrop can survive to a given height $z$ without totally evaporating, $r_\text{min}(z)$. We define $r_\text{min}(z)$ as the $r_0$ such that $r_\text{min}(z) - \Delta r$ evaporates before reaching $z$, but $r_\text{min}(z)$ reaches the $z$, i.e., $r_\text{eq}(z,r_0=r_\text{min})\geq \Delta r$. $r_\text{min}(z)$ is solved for via bisection by integrating $r(z)$ as described in section \ref{subsec:r_z} for initial radii between $\Delta r$ and the maximum raindrop radius described in section \ref{subsec:r_max}. 

On terrestrial planets (with a surface at $z_\text{surf}$), clouds that can grow raindrops of $r_0 \ge r_\text{min}(z_\text{surf})$ can move condensible mass from the atmosphere to the surface condensible reservoir. We can place an lower bound on raindrop size from the cloud-edge size threshold where a raindrop can survive to the surface without totally evaporating: $r_\text{min}(z_\text{surf})$, which we will henceforth abbreviate to simply $r_\text{min}$. 
On gaseous planets, there is no surface, and raindrops can only evaporate, but their ability to transport mass and heat as a function of height is still important dynamically.

\subsection{A Dimensionless Number Characterizing Raindrop Evaporation Regime}\label{subsec:nond_num}
To better understand raindrop evaporation, we simplify equation \eqref{eq:drdz} into a dimensionless number that can be more clearly interpreted---and evaluated---than a system of differential equations requiring numerical integration to solve. 
First, we need to simplify calculating $T_\text{drop}$ for an evaporating raindrop from the differential equation \eqref{eq:dTdt}.

We assume $T_\text{drop}$ changes only as a function of altitude. This is justified by comparing the timescale on which atmospheric temperature changes ($\tau_\text{air}$) to the timescale on which raindrop temperature changes ($\tau_\text{drop}$). Assuming a dry adiabatic temperature profile, $\tau_\text{air}\approx(c_{p\text{,air}}\Delta T_\text{air})/(g\text{d}z/\text{d}t)$ where we conservatively set the characteristic change in air temperature $\Delta T_\text{air}$ to 1 K. Assuming the atmosphere transfers heat to the raindrop via conduction, {$\tau_\text{drop}\approx (r_\text{eq}^2\rho_{\text{c,}\ell}c_{p\text{,c,}\ell})/(3 K_\text{air})$}. 
Except for the largest possible raindrops, under broad planetary conditions $\tau_\text{drop}\gg\tau_\text{air}$, and hence d$T_\text{drop}$/d$t$=0 is a good approximation at a given altitude.

We define $\Delta T_\text{drop}$ as the equilibrium temperature difference between the air and raindrop, i.e.,
\begin{linenomath*}
\begin{equation}
    \Delta T_\text{drop} \equiv T_\text{air}-\left.T_\text{drop}\right\rvert_{\text{d}T_\text{drop}/\text{d}t=0}.
\end{equation}
\end{linenomath*}
From this definition of $\Delta T_\text{drop}$ and equation \eqref{eq:dTdt},
\begin{linenomath*}
\begin{equation}\label{eq:equilibriumdelta}
    \Delta T_\text{drop} = \frac{D_\text{c-air} f_{V\text{,mol}}L\mu_\text{c}}{ K_\text{air} f_{V\text{,heat}}R}\left(\frac{p_\text{sat}(T_\text{air}-\Delta T_\text{drop})}{T_\text{air}-\Delta T_\text{drop}}-\text{RH}\frac{p_\text{sat}(T_\text{air})}{T_\text{air}} \right).
\end{equation}
\end{linenomath*}
This transcendental equation can be solved numerically via a root-finding algorithm. It is commonly simplified to an analytic expression using Clausius Clapeyron, Taylor expansions, and series of assumptions regarding $\Delta T_\text{drop}$ being small compared to $T_\text{air}$ \cite<e.g.,>[chapter 7]{Rogers1996}.

However, we find an analytic approximation that holds better across a broad range of planetary conditions is to evaluate equation \eqref{eq:equilibriumdelta} with $\Delta T_\text{drop}$ values on the right hand side approximated as
\begin{linenomath*}
\begin{equation}\label{eq:delta_simple}
   \Delta T_\text{drop} \approx 0.5(T_\text{air}-T_\text{LCL}). 
\end{equation}
\end{linenomath*}
$T_\text{drop}$ must fall between $T_\text{LCL}$ and $T_\text{air}$ (i.e, $\Delta T_\text{drop} \in [0,T_\text{air}-T_\text{LCL}]$) because there is no heat source for the drop once $T_\text{drop}=T_\text{air}$, and there is no heat sink for the drop once $T_\text{drop}=T_\text{LCL}$ because RH=1 and evaporation ceases. Equation \eqref{eq:delta_simple} can also be employed for a back-of-the-envelope calculation of $\Delta T_\text{drop}$.

Now we can define a dimensionless number $\Lambda$ to evaluate the tendency of a raindrop of radius $r_\text{eq}$ (and mass $m$) toward evaporation within a given vertical length scale $\ell$. $\Lambda$ is the ratio of evaporative mass loss during transit through $\ell$ to raindrop mass:
\begin{linenomath*}
\begin{equation}\label{eq:Lambda}
    \Lambda \equiv \frac{\ell}{m}\frac{\text{d}m}{\text{d}t}\left(\frac{\text{d}z}{\text{d}t}\right)^{-1} = \frac{3\ell}{r_\text{eq}}\frac{\text{d}r_\text{eq}}{\text{d}t}\left(\frac{\text{d}z}{\text{d}t}\right)^{-1}. 
\end{equation}
\end{linenomath*}
Here we have made use of the chain rule, the definition of $r_\text{eq}$, and the relation {$\text{d}m /\text{d}r_\text{eq}=4\pi \rho_{\text{c,}\ell}r_\text{eq}^2$}. Expanding the terms in $\Lambda$ gives
\begin{linenomath*}
\begin{equation}\label{eq:Lambda_expanded}
    \Lambda = \frac{3\ell}{r_\text{eq}^2}\frac{f_\text{V,mol} D_\text{c-air}\mu_\text{c}}{(w + v_T)\rho_{\text{c,}\ell}R}\left(\text{RH}\frac{p_\text{c,sat}(T_\text{air})}{T_\text{air}} - \frac{p_\text{c,sat}(T_\text{air}-\Delta T_\text{drop})}{T_\text{air}-\Delta T_\text{drop}}\right).
\end{equation}
\end{linenomath*}
Altitude-dependent values needed to calculate $\Lambda$ are evaluated at the midpoint of $\ell$. $\Delta T_\text{drop}$ can be evaluated from equation \eqref{eq:equilibriumdelta} numerically (most accurate), from equation \eqref{eq:equilibriumdelta} and \eqref{eq:delta_simple} algebraically, or from equation \eqref{eq:delta_simple} (back-of-envelope). $v_\text{T}$ can be evaluated from equation \eqref{eq:v} numerically or from a parameterized relationship for a commonly studied planet. Alternatively, $v_\text{T}$ can be estimated via Stokes law for very small drops or via $v_\text{T,max}\approx2\left(\sigma_\text{c-air}(\rho_{\ell\text{,c}}-\rho_\text{air})g\right)^{0.25}\left(\rho_\text{air}\right)^{-0.5}$ for very large drops \cite<>[chapter 7.C]{Clift2005}.

$\Lambda$ values give the expected change in raindrop mass from evaporation relative to initial mass after falling a given distance. $\Lambda(r_\text{eq},\ell)\ge 1$ indicates raindrops of size $r_\text{eq}$ will fully evaporate over distance $\ell$. 
 Therefore, the fraction of raindrop mass evaporated over $\ell$ can be estimated from min\{$\Lambda$,1\}. For a given $\ell$, the $r_\text{eq}$ such that $\Lambda=1$ approximates the minimum radius to reach that distance below the starting $z$ without fully evaporating, $r_\text{min}(z_\text{start}-\ell)$. 
 For simplicity, here we only consider $\Lambda$ defined when d$z$/d$t<$0, i.e., when a raindrop is falling downward. Though we do not treat raindrop formation here, we note that with some slight modifications this dimensionless number can also be employed to consider the effectiveness of cloud drop growth via condensation.

\subsection{Maximum Raindrop Size Before Breakup}\label{subsec:r_max}
The final physical process we need to consider is raindrop breakup. Raindrops cannot grow to infinitely large sizes because the resistance provided by surface tension as surface area increases is limited. When surface tension ceases to be the dominant force experienced by a raindrop, the raindrop rapidly breaks apart.

A variety of approaches to estimating this maximum stable raindrop radius $r_\text{max}$ have been proposed previously; but none are expected to yield quantitatively exact values. The physics is additionally complicated in many situations (such as on present-day Earth) by the fact that the practical upper bound on raindrop size is not set from individual raindrop breakup but rather from hydrometeor collisions \cite<e.g.,>[]{Barros2010}. 
Given the uncertainties, we are therefore primarily concerned here with how $r_\text{max}$ scales with external planetary properties. In particular, we focus on the effect of air density, which has inconsistently been claimed within the planetary literature to have no effect on $r_\text{max}$ \cite{Som2012,Palumbo2020} and an extremely significant one \cite{Craddock2017}.

Variants of two methods have commonly been used to describe raindrop breakup; we review them here and describe their origins in more detail in \ref{app:breakup}.
First, we can estimate $r_\text{max}$ by considering when the base of a raindrop becomes unstable to small perturbations from a more dense fluid (liquid condensible) being on top of a less dense fluid (air)---generally referred to as Rayleigh-Taylor instability \cite<>[chapter 10.3.4]{Komabayasi1964,Grace1978,Lehrer1975,Clift2005,Pruppacher2010}. This analysis yields a maximum length scale $\ell_\text{RT,max}$ that can be related back to a maximum equivalent radius $r_\text{max}$:
\begin{linenomath*}
\begin{equation}\label{eq:ell_max_rt}
    \ell_\text{RT,max} = \pi \sqrt{\frac{\sigma_{\text{c-air}}}{g(\rho_{\text{c},\ell} - \rho_\text{air})}}.
\end{equation}
\end{linenomath*}
There is not a definitive $\ell_\text{RT,max}$, with different authors choosing related, but often distinct, length scales.

Another common approach for estimating $r_\text{max}$ in the Earth literature is to calculate when the force of surface tension $F_\sigma$ is balanced by the aerodynamic drag force \cite<e.g.,>[chapter 10.3.4]{Pruppacher2010}. We henceforth refer to this approach as ``force balance.'' Again, we get a relationship to be solved for $r_\text{max}$ that depends on a somewhat arbitrary length scale, here pertaining to surface tension $\ell_{\sigma\text{,max}}$:
\begin{linenomath*}
\begin{equation}\label{eq:ell_max_fb}
    \frac{r_\text{max}^3}{\ell_{\sigma\text{,max}}} = \frac{3}{4\pi}\frac{\sigma_\text{c-air}}{(\rho_{\text{c,}\ell}-\rho_\text{air})g}.
\end{equation}
\end{linenomath*}

We evaluate equations \eqref{eq:ell_max_rt} and \eqref{eq:ell_max_fb} for $r_\text{max}$ under different length scales proposed in the literature. Length scales are related to $r_\text{max}$ via the geometry of spheres or oblate spheroids. Both approaches yield similar expressions for $r_\text{max}$ with some variation in dependence on raindrop shape and constant factors depending on the choice of length scale.

\section{Results}\label{sec:results}
Having described the key physical processes that affect isolated falling raindrops in detail, we now present numerical results for a wide range of planetary conditions and circumstances pertaining to falling raindrops.
We validated our shape and terminal velocity calculations against modern Earth observations, experimental results, and empirically based calculations \cite<Figures S3-S4;>[]{Gunn1949,Best1950,Pruppacher1970,Pruppacher1971,Beard1976,Beard1987,Thurai2009}. We also compared, with reasonable agreement, our results to previous planetary theoretical results on Titan's methane-nitrogen raindrops for shape, terminal velocity, and raindrop properties with altitude \cite<Figures S5-S8; Table S1;>[]{Lorenz1993,Graves2008}, using Cassini Huygens' probe data where appropriate \cite{Fulchignoni2005,Niemann2005}.

\begin{sidewaystable}
\caption{\textit{Planetary properties used in calculations}}
\begin{tabular}{|l||l|l|l|l|l|l|l|l|l|l||l|}
\hline
planet     & $z_\text{ref}$ & $T(z_\text{ref})$     & $p_\text{dry}(z_\text{ref})$         & RH$(z_\text{ref})$   & $g$   & $f_\text{\ce{H2},dry}$ & $f_\text{\ce{He},dry}$ & $f_\text{\ce{N2},dry}$ & $f_\text{\ce{O2},dry}$ & $f_\text{\ce{CO2},dry}$  & $H_\mathrm{LCL}$ \\
name       &           & [K]     & [$10^5$ Pa] & [ ]  & [m s$^{-1}$]  & [mol mol$^{-1}$]       & [mol mol$^{-1}$]       & [mol mol$^{-1}$]       & [mol mol$^{-1}$]       & [mol mol$^{-1}$] & [km]        \\
\hline
\hline
Earth-like$^{a}$ & surface   & 300     & 1.01325     & 0.75 & 9.82   & 0                      & 0                      & 1                      & 0                      & 0     &     8.97                     \\
\hline
Earth      & surface   & 290     & 1.01325     & 0.75 & 9.82       & 0                      & 0                      & 0.8                    & 0.2                    & 0   & 8.41                       \\
\hline
early Mars$^{b}$       & surface   & 290     & 2           & 0.75 &   3.71         & 0                      & 0                      & 0                      & 0                      & 1    & 14.5                    \\
\hline
Jupiter    & LCL       & 274$^{c}$     & 4.85$^{c}$        & 1    &   24.84   & 0.864$^{d}$       &0.136$^{d}$           & 0                      & 0                      & 0       &   39.8                 \\
\hline
Saturn     & LCL       & 284$^{c}$     & 10.4$^{c}$        & 1    & 10.47    &    0.88$^{d}$    &       0.12$^{d}$             & 0                      & 0                      & 0   &  99.2                    \\
\hline
K2-18b$^{e}$     & LCL       & 275  & 0.1         & 1    & 12.44$^{f,g}$       & 0.9                    & 0.1                    & 0                      & 0                      & 0  &  56.6                       \\
\hline
composition & LCL  & 275  & 0.75         & 1    & 9.82       & 0,1                     & 0,1                     & 0,1                        & 0,1                       & 0,1          & 5.32-108             \\ 
\hline
broad      & LCL       & 275-400 & 0.05-100    & 1    & 2-25          & 0,1                    & 0                      & 0,1                    & 0                      & 0,1    & 2.09-528     \\
\hline
\end{tabular}
\caption*{\textit{Note.} Input properties (columns between the double vertical lines) are specified for reference altitude $z_\text{ref}$ and used to determine atmospheric properties under the cloud layer following the assumptions outlined in section \ref{subsec:r_z}. $p_\text{dry}$ is the pressure of all non-condensible gas species. $f$ is the dry molar concentration. Atmospheric scale height $H\equiv RT(g\mu_\text{avg})^{-1}$ is evaluated at $z_\text{LCL}$ throughout this paper. Vertical wind speed is set to 0 m s$^{-1}$ unless otherwise specified in the text. $^{a}$ $T_\text{surf}$ is higher than average Earth $T_\text{surf}$ in order to highlight a larger range of possible surface RH values while keeping $T_\text{LCL}$ above freezing. Pure \ce{N2} background atmosphere is assumed for simplicity. $^{b}$Speculative values for hypothesized warm, wet period in Mars' ancient past. $^{c}$\citeA{Carlson1988}. $^{d}$\citeA{Leconte2017}. $^{e}$ Speculative values compatible with constraints from \citeA{Benneke2019}. $^{f}$\citeA{Cloutier2019}. $^{g}$\citeA{Benneke2019}. } 
\label{tab:conditions}
\end{sidewaystable}

\subsection{Raindrop Evaporation under Earth-like Conditions}\label{subsec:res_evap}
\begin{figure}
\centering
\includegraphics[width=0.9\textwidth]{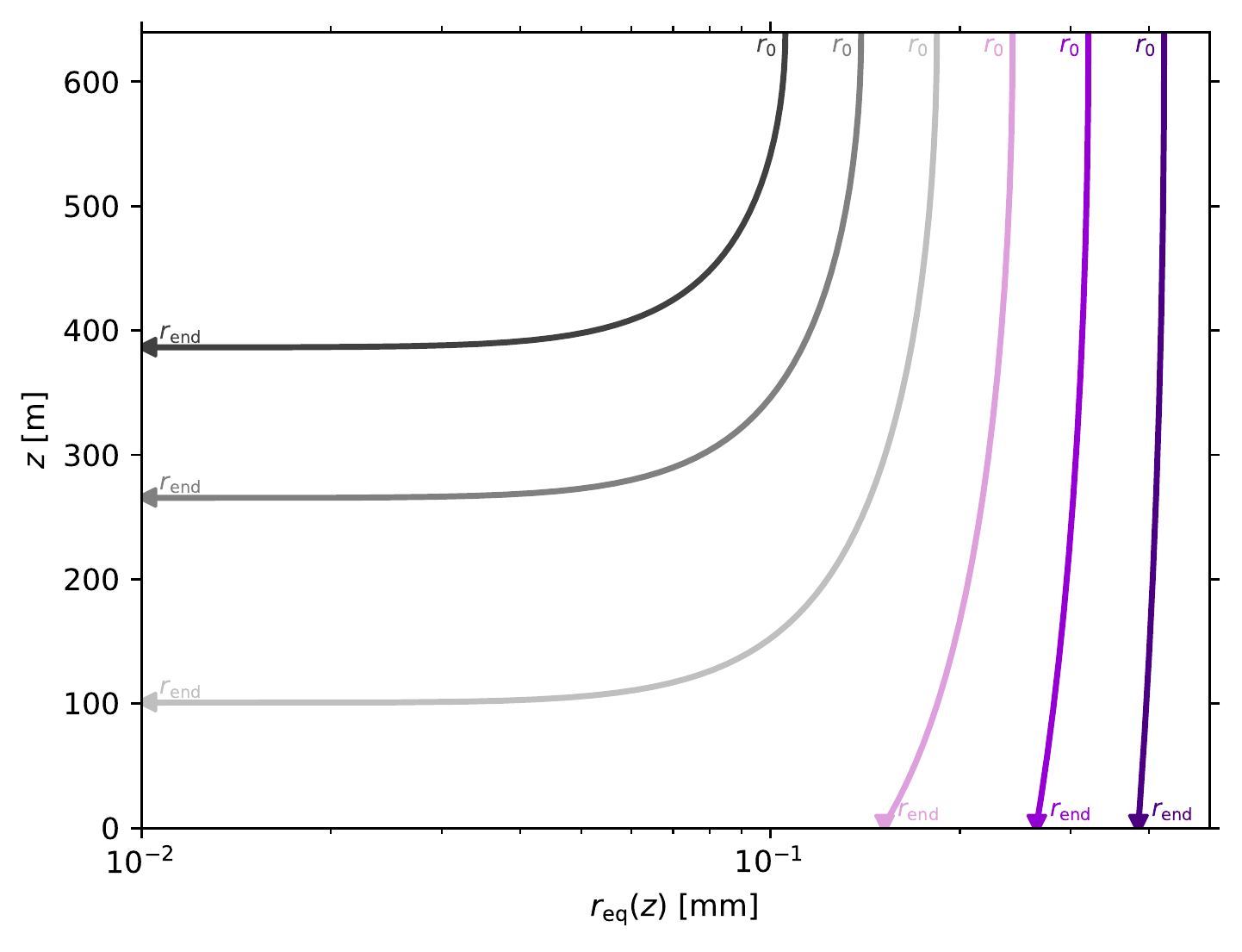}
\caption{Raindrop altitude $z$ versus equivalent radius $r_\text{eq}$ for equally log-spaced initial raindrop radii ($r_0$) near the minimum radius threshold for survival to surface ($r_\text{min}$). Gray-shaded lines are raindrops that evaporate before reaching the surface while purple-shaded lines are raindrops that successfully reach the surface. Planetary conditions are set to Earth-like as given in Table \ref{tab:conditions}.}
\label{fig:r_z_many_r0s}
\end{figure}

Integrating equation \eqref{eq:drdz}, we investigated the behavior of raindrop evaporation for water raindrops falling from the cloud base to planetary surface under Earth-like conditions (Table \ref{tab:conditions}). Figure \ref{fig:r_z_many_r0s} shows the evolution of raindrop radius as function of altitude $z$ for a number of an initial radii at cloud base until the raindrops either completely evaporate ($r_0 < r_\text{min}$) or reach the surface ($r_0 \geq r_\text{min}$). The results imply a strong positive feedback on raindrop evaporation as raindrops grow smaller. 

\begin{figure}
\centering
\includegraphics[width=\textwidth]{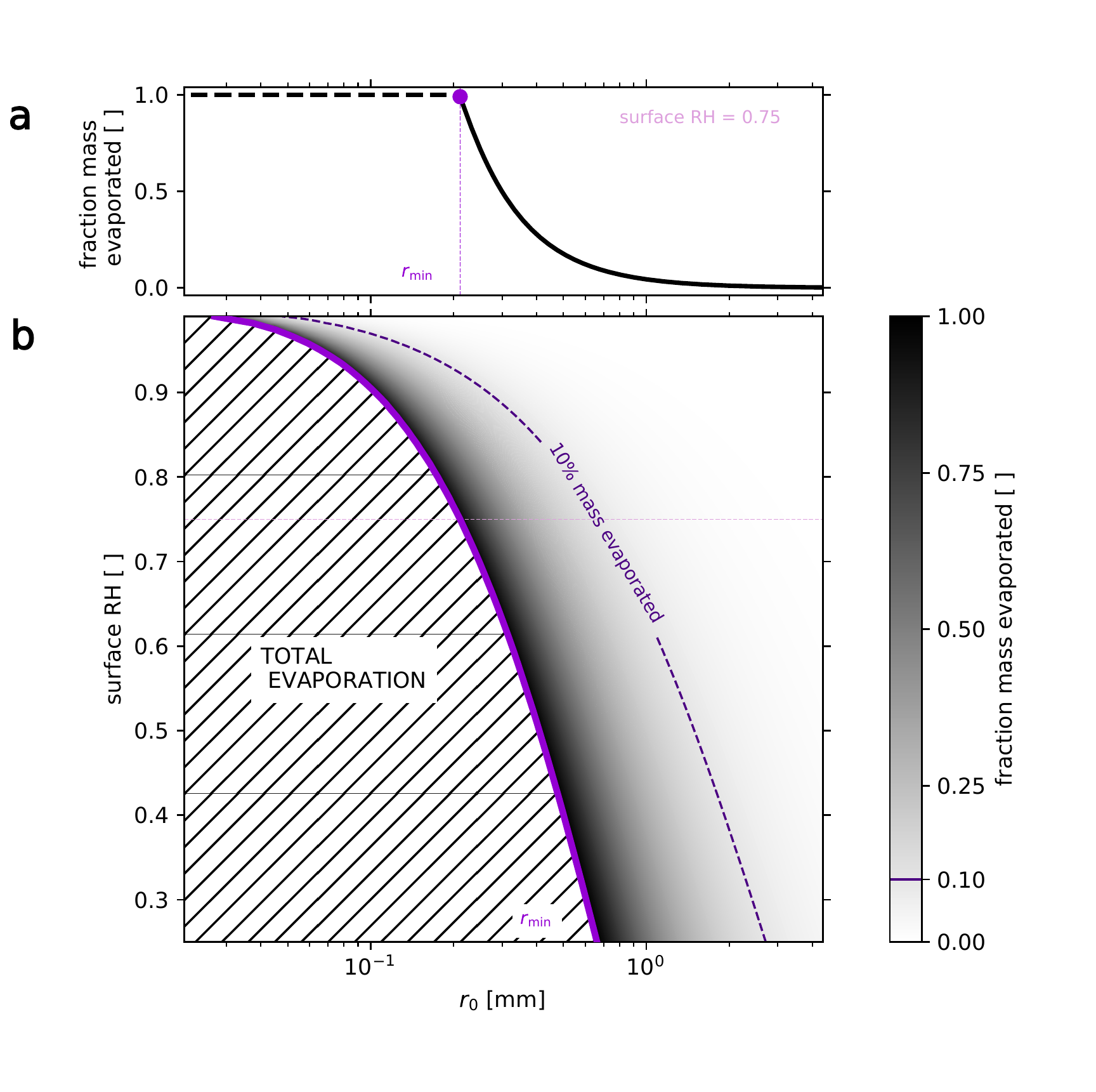}
\caption{(a) Fraction of raindrop mass evaporated at the surface versus initial radius $r_0$ for an Earth-like planet (Table \ref{tab:conditions}). The purple dot marks $r_\text{min}$, the $r_0$-threshold for a raindrop to reach the surface without totally evaporating. (b) Fraction of raindrop mass evaporated (black-white color scale) versus surface relative humidity and $r_0$. The same as the top panel except with varying surface RH; the horizontal light-purple line highlights the surface RH slice that the top panel displays. The purple line marks the calculated $r_\text{min}$ as a function of surface RH. For $r_0 < r_\text{min}$, raindrops totally evaporate before reaching the surface (hatched region). The dashed dark-purple line highlights the 10\% mass evaporated contour within the more continuous shading.}
\label{fig:frac_mass_evap_RH}
\end{figure}

Figure \ref{fig:frac_mass_evap_RH}(a) demonstrates this positive feedback more explicitly. For the same planetary conditions as in Figure \ref{fig:r_z_many_r0s}, it shows the fraction of raindrop mass evaporated at the surface for a range of $r_0$ values. This curve approaches a step function about $r_\text{min}$.  Figure \ref{fig:frac_mass_evap_RH}(b) extends \ref{fig:frac_mass_evap_RH}(a) by showing fraction of raindrop mass evaporated via colormap versus surface relative humidity and $r_0$. Qualitatively, Figure \ref{fig:frac_mass_evap_RH}(b) shows the same sharp cut-off behavior as \ref{fig:frac_mass_evap_RH}(a). 
Surface RH affects the quantitative value of $r_\text{min}$ because it varies RH($z$), which impacts the magnitude of evaporation rate as well as the height of $z_\text{LCL}$---a higher surface RH gives a lower $z_\text{LCL}$. Both these effects act to make $r_\text{min}$ decrease as surface RH increases. 

There is not yet an analytical method for estimating average surface RH in a generic planetary atmosphere---in part because of the poorly understood feedbacks of precipitation evaporation on average  RH \cite{Romps2014,Lutsko2018}---so, for now, we consider surface RH a prescribed planetary parameter. In this plot, we vary surface RH from 99.9\% to 25\%---the former arbitrarily close to the threshold for evaporation to begin (RH < 100\%) and the latter about the minimum surface RH for which the temperature at $z_\text{LCL}$ is above the freezing point of \ce{H2O} given our other chosen planetary parameter values. 

\begin{figure}
\centering
\includegraphics[width=\textwidth]{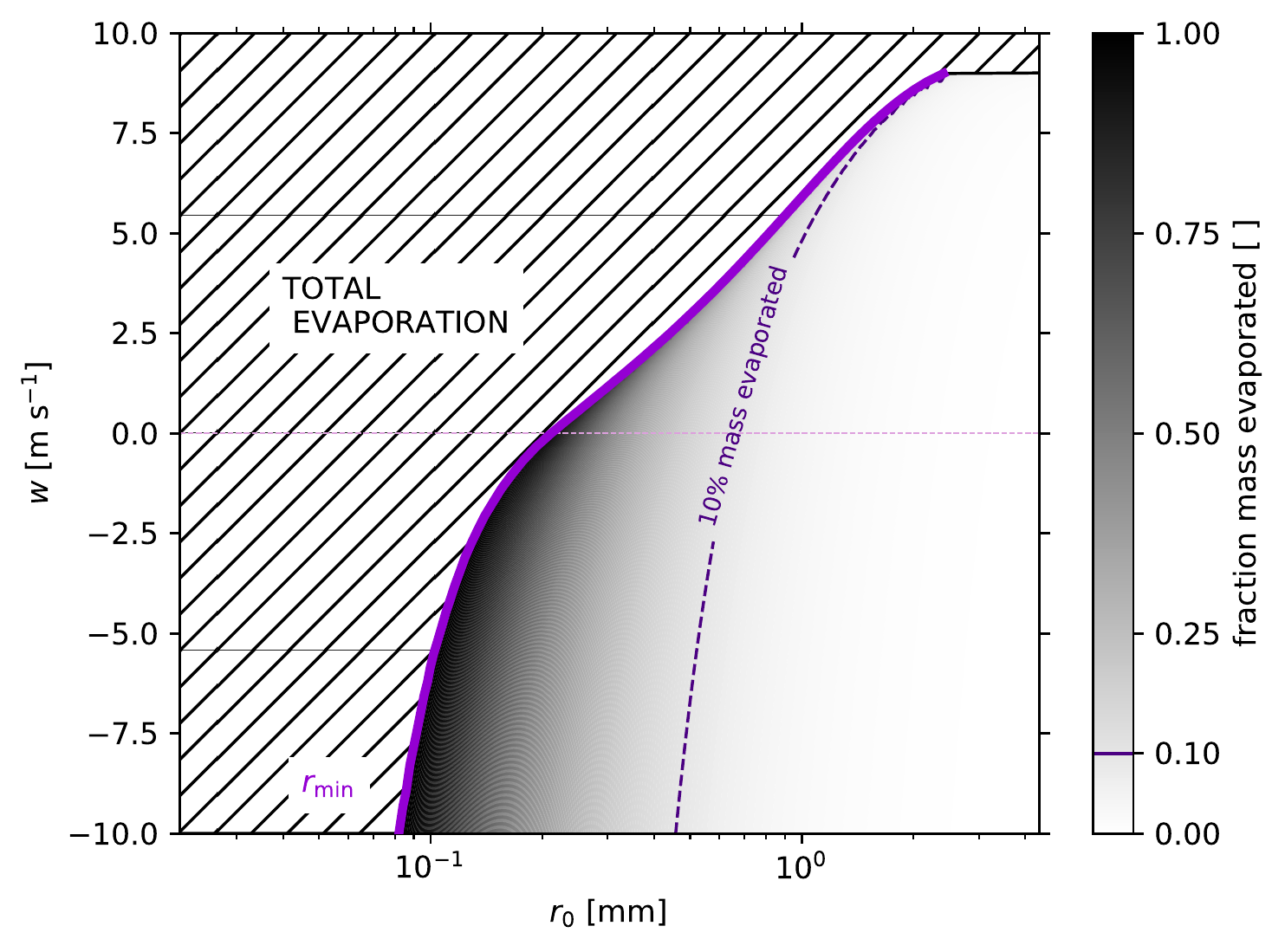}
\caption{Fraction of raindrop mass evaporated (black-white color scale) versus vertical wind speed $w$ and initial radius $r_0$ for an Earth-like planet (Table \ref{tab:conditions}). The horizontal light-purple line ($w=0$) divides updrafts ($w > 0$) from downdrafts ($w<0$). The purple line marks the calculated $r_\text{min}$ as a function of $w$. For $r_0 < r_\text{min}$, raindrops totally evaporate before reaching the surface (hatched region). The dashed dark-purple line highlights the 10\% mass evaporated contour within the more continuous shading.}
\label{fig:frac_mass_evap_w}
\end{figure}

Figure \ref{fig:frac_mass_evap_w} is the same as Figure \ref{fig:frac_mass_evap_RH}(b) except it probes the effect of vertical wind speed $w$ rather than surface RH. Downdrafts ($w < 0$) increase the falling speed of raindrops while updrafts ($w > 0$) decrease the falling speed of raindrops (as long as $w + v_\text{T} < 0$). Updrafts can transport raindrops upward once the updraft speed exceeds the magnitude of a raindrop's terminal velocity. 
As downdraft speed increases, there is a smoother transition through fraction mass evaporated across $r_0$ values. As updraft speed increases, fraction mass evaporated approaches a step function about $r_\text{min}$ until $w + v_\text{T,max} = 0$. For updrafts speeds greater than this threshold, no $r_\text{min}$ exists as raindrops are no longer falling. 

As with surface relative humidity, there is no analytic approach for estimating average $w$ ranges in generic planetary atmospheres (though values can be probed by mesoscale models of sufficient resolution).
In this plot, we bound updraft speed from where $w$ exceeds maximum raindrop terminal velocity and then choose a symmetric downdraft speed bound. (This choice of a lower bound is arbitrary and does not represent an end-member case for downdraft speeds.) For simplicity, here we fix $w$ as constant throughout raindrop falling and evaporation. In reality, vertical velocities vary both spatially and temporally within a given storm event \cite<e.g.,>[]{Lohmann2016}, often as a result of the interaction between precipitation particles and ambient air \cite<e.g.,>[]{Rogers1996}. 

\subsection{Evaluation of Dimensionless Number Characterizing Raindrop Evaporation Regime}\label{subsec:results_Lambda}
\begin{figure}
\centering
\includegraphics[width=\textwidth]{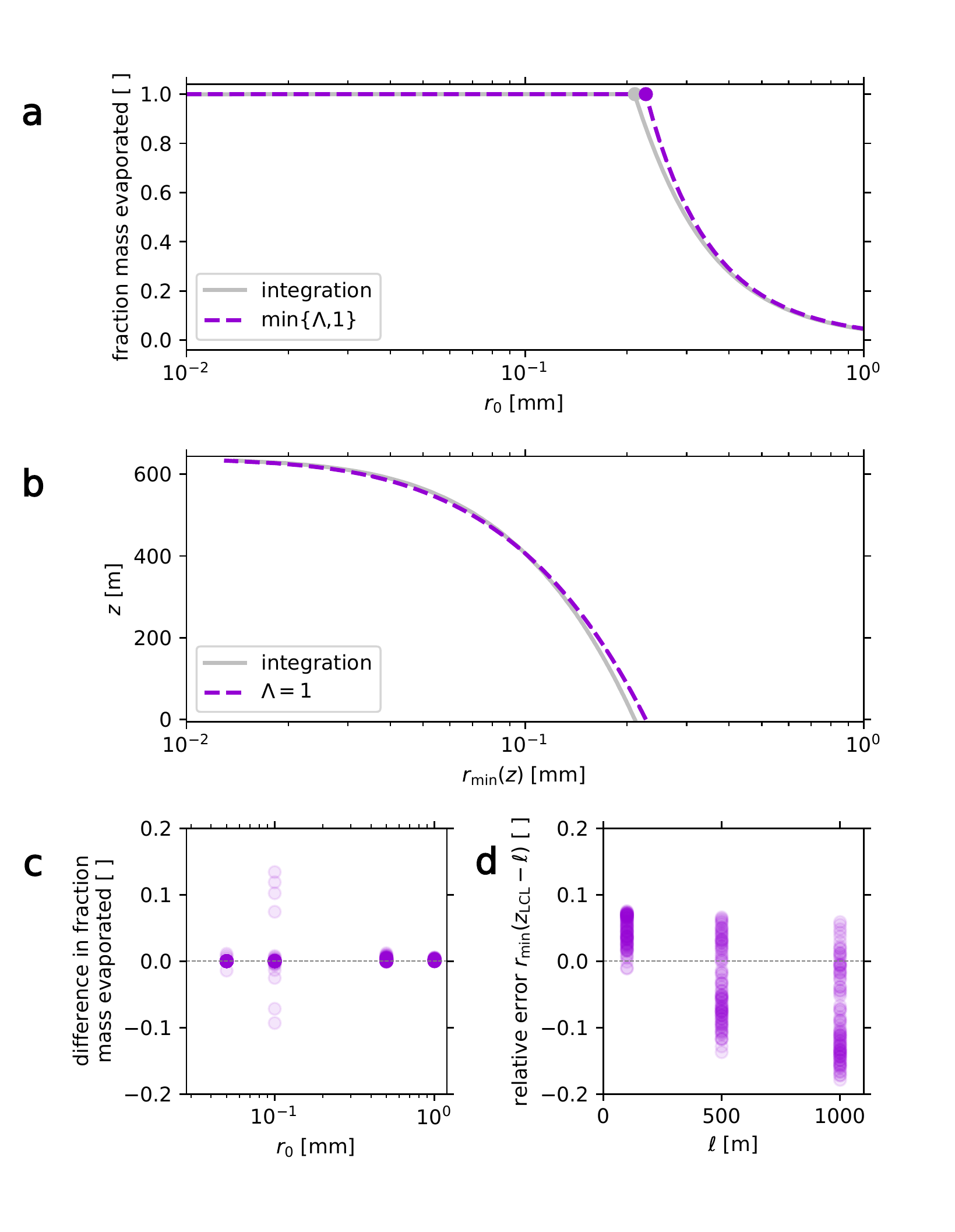}
\caption{(a) Fraction mass evaporated at the surface versus initial raindrop radius $r_0$ for Earth-like atmospheric conditions (Table \ref{tab:conditions}) evaluated from numerical integration (dashed gray line) and using dimensionless number $\Lambda$ (purple line). (b) Altitude $z$ versus threshold initial raindrop radius for total evaporation at $z$ ($r_\text{min}(z)$) for Earth-like atmospheric conditions evaluated from numerical integration (dashed gray line) and using dimensionless number $\Lambda$ (purple line).  (c) Difference in fraction mass evaporated between calculations using numerical integration and $\Lambda$ versus four $r_0$ values evaluated across broad planetary conditions (Table \ref{tab:conditions}) at 500 m below cloud base. (d) Relative error in $r_\text{min}(z)$ calculated using $\Lambda$ relative to numerical integration versus three $\ell$ values across broad planetary conditions. For (c) and (d), scatter points are semi-transparent to highlight where points cluster.}
\label{fig:nond_v_num_int}
\end{figure}

Figure \ref{fig:nond_v_num_int} compares using the dimensionless number $\Lambda$ defined in equation \eqref{eq:Lambda} to predict raindrop evaporation behavior to using numerical integration for Earth-like atmospheric conditions (a)-(b) and across broad planetary conditions (c)-(d).  Figure \ref{fig:nond_v_num_int}(a) shows calculations of the fraction of raindrop mass evaporated at the surface relative to initial mass at cloud base versus initial radius $r_0$ using both numerical integration and the minimum of $\Lambda(r_0,\ell=z_\text{LCL})$ and 1. Figure \ref{fig:nond_v_num_int}(c) is similar to (a) except it shows the difference in fraction mass evaporated at 500 m below the cloud base between these two methods at 4 $r_0$ for 90 different planetary conditions. The ``broad'' conditions in Table \ref{tab:conditions} give the ranges over which we vary $T$, $p$, and $g$ at cloud base for background gas atmospheres of pure \ce{H2}, \ce{N2}, and \ce{CO2}. Only one value among $T$, $p$, and $g$ is changed at a time relative to the ``composition'' conditions, which are used as a baseline. 

Figure \ref{fig:nond_v_num_int}(b) shows calculations of the threshold minimum radius to reach altitude $z$ without fully evaporating ($r_\text{min}(z)$) using both numerical integration and $\Lambda$=1.
Figure \ref{fig:nond_v_num_int}(d) is similar to (b) except it shows the relative error in $r_\text{min}(z)$ calculated using $\Lambda$ instead of numerical integration at three $\ell$ values for the same 90 different planetary conditions as described for panel (c).

Unsurprisingly, we find that the accuracy of using $\Lambda$ to calculate the fraction of raindrop mass evaporated at $z$ decreases for $r_0$ near $r_\text{min}(z)$ and that the accuracy of using $\Lambda$ to calculate $r_\text{min}(z)$ decreases as the distance between the cloud base and $z$ increases. Nonetheless, Figure \ref{fig:nond_v_num_int} demonstrates that $\Lambda$ can capture the essential behavior of fraction mass evaporated and $r_\text{min}(z)$ with a small fraction of the computational cost of numerical integration ($\ll 1\%$). Comparing the use of $\Lambda$ and a full numerical integration for calculations at $\ell=500$ m, we found percent errors for $r_\text{min}(z)$ and percent differences in fraction raindrop mass evaporated were usually less than $10\%$ in magnitude across a broad planetary parameter space, and even the largest errors were less than 20\%. 
This agreement indicates that $\Lambda$ is a viable way to predict and interpret raindrop evaporation regimes. In \ref{app:nond}, we use the definition of $\Lambda$ to show the numerical results of Figures \ref{fig:r_z_many_r0s} and \ref{fig:frac_mass_evap_RH} can be understood from an analytic mathematical perspective.

\subsection{Investigation into the Dependence of Maximum Raindrop Size on Air Density}

\begin{figure}
\centering
\includegraphics[width=\textwidth]{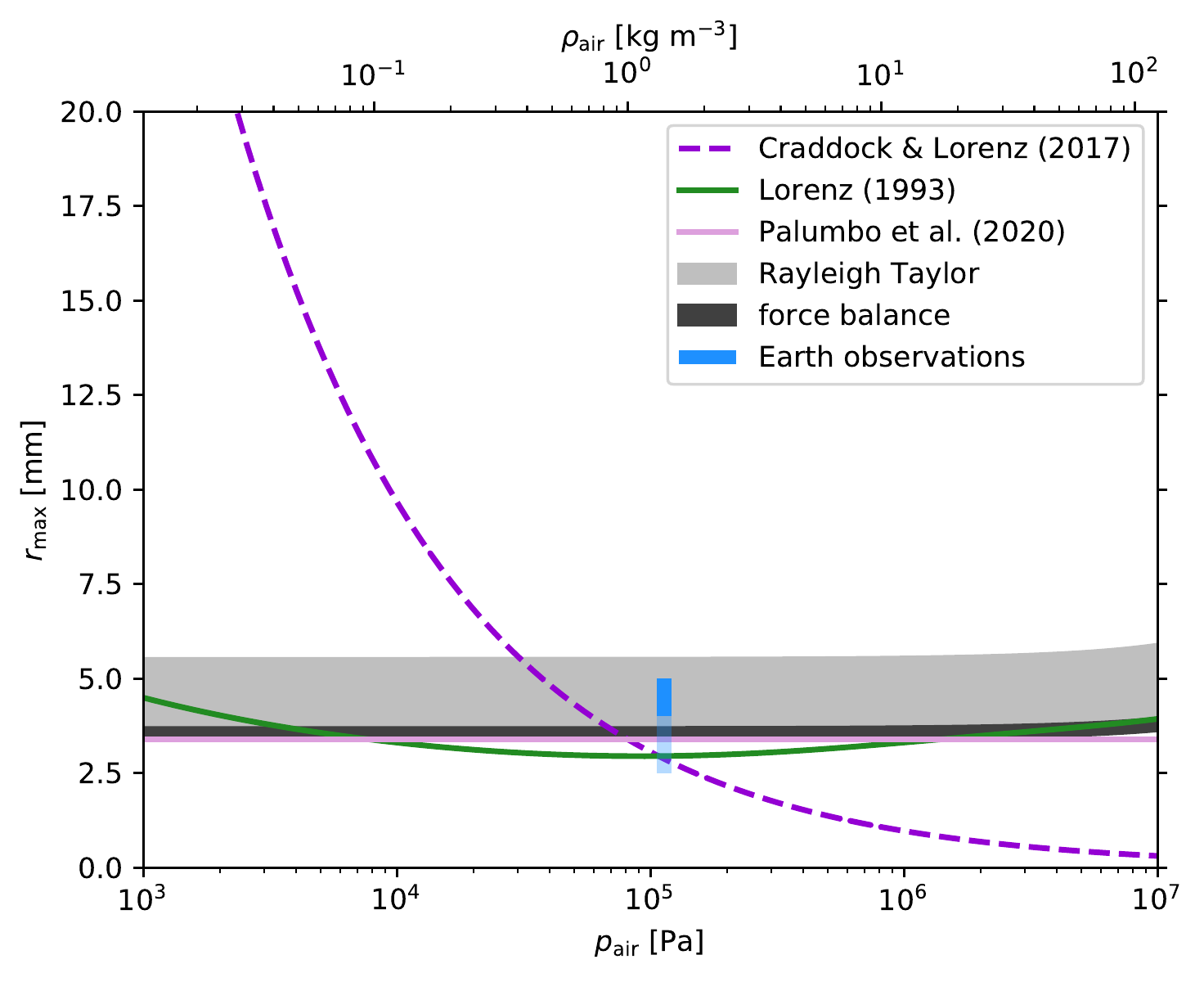}
\caption{Maximum stable water raindrop size $r_\text{max}$ versus air pressure $p_\text{air}$ or air density $\rho_\text{air}$, assuming fixed temperature {$T= 275$ K} and {$\text{RH}=0$} (for a linear relationship between $p_\text{air}$ and $\rho_\text{air}$), with Earth surface gravity and \ce{N2} background gas. Different line colors correspond to different methods for calculating $r_\text{max}$ as labeled in the legend and described in the text. The Rayleigh-Taylor and force balance methods can use (arbitrarily) different length scales, so we plot the span of values from common length scale choices rather than single lines.}
\label{fig:r_max_v_pressure}
\end{figure}

Before we combine the concepts of minimum and maximum raindrop size thresholds, we return to the debate in the planetary literature over whether maximum stable raindrop size $r_\text{max}$ depends on air density. In Figure \ref{fig:r_max_v_pressure}, we compare $r_\text{max}$ as a function of air density (or air pressure) using break up criteria as presented in section \ref{subsec:r_max} and previous planetary literature. The two approaches we reviewed---Rayleigh-Taylor instability and force balance---are dependent on somewhat arbitrary length scales, so we plot the spread in values from different length scales. For Rayleigh-Taylor instability, we consider $\ell_\text{RT,max}=0.5\pi r_\mathrm{eq},0.5\pi a,2r_\mathrm{eq},2a$; for force balance we consider $\ell_{\sigma\text{,max}} = 2\pi r_\text{eq}, 2\pi a$.

From the planetary literature on maximum raindrop size \cite{Rossow1978,Lorenz1993,Som2012,Craddock2017,Palumbo2020}, we plot in Figure \ref{fig:r_max_v_pressure} the approaches of \citeA{Lorenz1993}, \citeA{Craddock2017}, and \citeA{Palumbo2020}, which we will describe in more detail shortly.
We do not plot \citeA{Som2012}'s quantitative approach as they make use of empirical fits and Earth-based observations, but in practice they suggest a similar criterion to \citeA{Lorenz1993}. \citeA{Rossow1978} and the \citeA{Clift2005} expression cited in \citeA{Lorenz1993} use Rayleigh-Taylor instability criteria with length scales included in our range.
 
To calculate $r_\text{max}$, \citeA{Lorenz1993}, \citeA{Craddock2017}, and \citeA{Palumbo2020} all begin from the same criterion: the raindrop radius where the dimensionless Weber number equals 4. The Weber number We, which characterizes the ratio of drag force to the force of surface tension, is defined as
\begin{linenomath*}
\begin{equation}\label{eq:We}
    \text{We} \equiv \frac{r_\text{eq}v^2\rho_\text{air}}{\sigma_\text{c-air}}.
\end{equation}
\end{linenomath*}
This approach is equivalent to the force balance approach under the assumption that, within $F_\text{drag}$, cross sectional area times $C_\text{D}$ is equal to $\pi r_\text{eq}^2$---an assumption that holds reasonably well under Earth surface conditions \cite{Matthews1964}.

To calculate $r_\text{max}$, \citeA{Lorenz1993} solves for the $r_\text{eq}$ satisfying {We = 4} numerically (to account for the dependence of $v$ on $r_\text{eq}$). As seen in Figure \ref{fig:r_max_v_pressure}, this setup yields an $r_\text{max}$ that varies by about a factor of 1.5 over the range of $\rho_\text{air}$ we consider. We view this variation with density as a non-physical dependence introduced by the neglect of the dimensionless $C_\text{D}$ in representing the drag force in the formulation of We \cite<>[chapter 8]{Kolev2007}. ($C_\text{D}$ nonlinearly depends on $\rho_\text{air}$ through Re.) Varying the $C_\text{D}$ parameterization within this calculation causes comparable changes in $r_\text{max}$ to varying $\rho_\text{air}$; thus, we consider the \citeA{Lorenz1993} method consistent with no significant dependence of $r_\text{max}$ on $\rho_\text{air}$. To calculate $r_\text{max}$, \citeA{Palumbo2020} solves for the $r_\text{eq}$ satisfying {We = 4} algebraically after assuming spherical raindrops and $C_\text{D}$=1:
\begin{linenomath*}
\begin{equation}\label{eq:r_maxP20}
    r_\text{max} = \left(\frac{3\sigma}{2g\rho_{\text{c,}\ell}}\right)^{0.5}.
\end{equation}
\end{linenomath*}

\citeA{Craddock2017} does not present a simplified expression for calculating $r_\text{max}$, only evaluations under different atmospheric conditions and a statement that ``larger diameter raindrops are not possible at higher atmospheric pressures.'' % caption to figure 2
When we simplify their presented equations involved in describing $r_\text{max}$ (their equations (1) and (3)) following their stated assumptions, we arrive at the same $r_\text{max}$ result as Equation \eqref{eq:r_maxP20}.
We are only able to reproduce the results of \citeA{Craddock2017}'s evaluations of $r_\text{max}$ (their Table 1) using an expression for $r_\text{max}$ inconsistent with length units:
\begin{linenomath*}
\begin{equation}
    r_\text{max} = \left(\frac{3\sigma}{2g\rho_\text{air}\rho_{\text{c,}\ell}}\right)^{0.5}.
\end{equation}
\end{linenomath*}

In addition to theoretical methods of estimating $r_\text{max}$, we also plot in Figure \ref{fig:r_max_v_pressure} a range of claimed maximum raindrop sizes on present-day Earth, both from experiments and natural observations. This range is consistent in magnitude with all the estimates. We give a range of maximum values as the measurement is an attempt to estimate the end of the extreme tail of a stochastic process \cite<e.g.,>[]{Komabayasi1964,Grace1978,Clift2005}. Single-value maxima fall between $r_\text{eq}$ of 4-5 mm \cite{Merrington1947,Beard1969,Ryan1976,Hobbs2004,Gatlin2015}. \citeA{Gatlin2015} compiled observations of $2.4\times10^8$ raindrops and found 0.4\% of these raindrops had $r_\text{eq}\ge2.5$ mm and only $1.9\times10^{-5}\%$ had $r_\text{eq}\ge4$ mm---statistics that suggest in practice $r_\text{max}$ is about 2.5-4 mm. 

Returning to theory, in Figure \ref{fig:r_max_v_pressure} we see that while different assumptions about raindrop shapes in the Rayleigh-Taylor and force balance methods lead to factor of a few quantitative differences in $r_\text{max}$, these differences are not sensitive to air density. We conclude that the effects of raindrop shape are ultimately of limited importance in estimating $r_\text{max}$ due to the ambiguity of raindrop length scales in the calculation setups. Thus the scalings of $r_\text{max}$ with planetary variables can reliably be seen analytically by assuming spherical raindrops (i.e., setting $b$/$a$ = 1):
\begin{linenomath*}
\begin{equation}
    r_\text{max} \propto \sqrt{\frac{\sigma_\text{c-air}}{g(\rho_{\text{c},\ell} - \rho_\text{air})}} \approx \sqrt{\frac{\sigma_\text{c-air}}{g\rho_{\text{c},\ell}}}.
\end{equation}
\end{linenomath*}
The force balance, Rayleigh-Taylor instability, and Weber number methods all yield the same approximate scalings, which are effectively independent of air density. (These scalings are not novel \cite<e.g.,>[]{Clift2005}; we simply present them in the context of this planetary debate on $r_\text{max}$.)

Therefore, we agree with \citeA{Palumbo2020} that \citeA{Craddock2017}'s finding that larger raindrops become possible as time advances and Mars experiences atmospheric escape due the dependence of raindrop stability on $\rho_\text{air}$ is not justified.  
This conclusion is also consistent with extensive modern-Earth based literature considering $r_\text{max}$, which does not explicitly highlight any dependence of $r_\text{max}$ on air density, which non-trivially varies from cloud to surface. 
Finally, we note that we are not claiming that average raindrop size is insensitive to air density, only that the instability of individual large raindrops does not have a significant dependence on air density. 

\subsection{Raindrop Size Bounds for Terrestrial Planets}
\begin{figure}
\centering
\includegraphics[width=\textwidth]{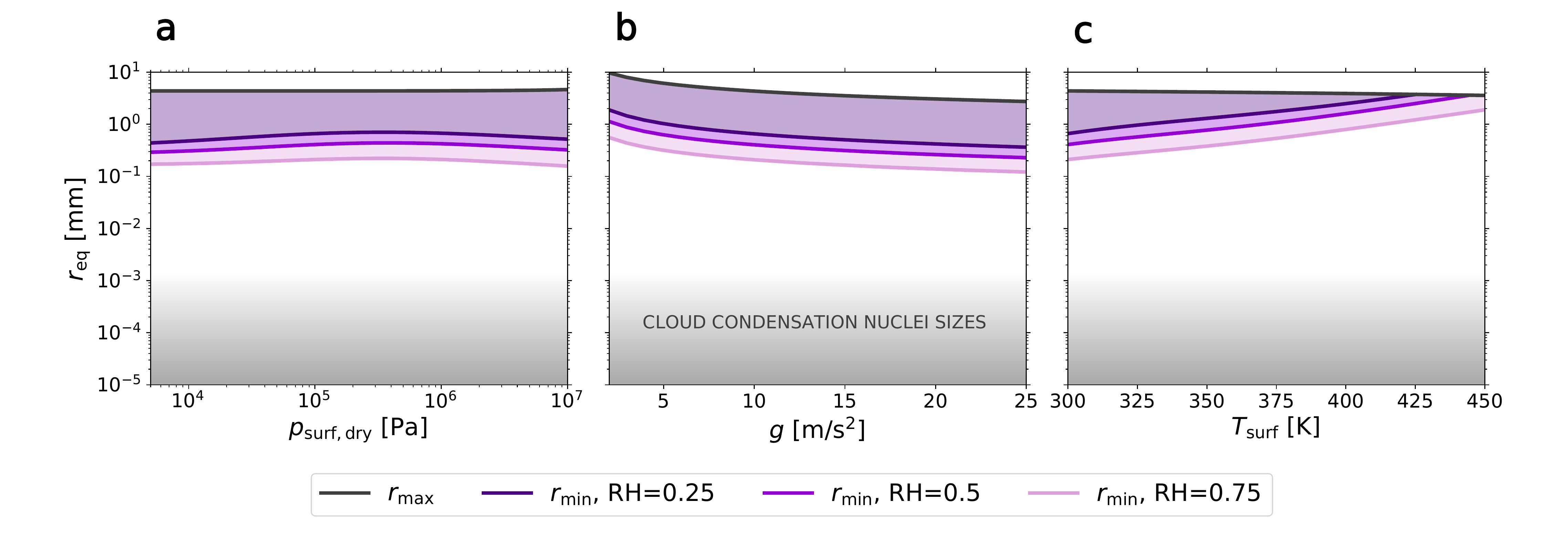}
\caption{Threshold $r_\text{eq}$ values versus (a) dry planetary surface pressure $p_\text{surf,dry}$, (b) surface gravity $g$, and (c) surface temperature $T_\text{surf}$. Planetary conditions not explicitly varied follow the Earth-like conditions in Table \ref{tab:conditions}. Dark-gray lines give $r_\text{eq}$ at the onset of Rayleigh-Taylor instability ($r_\text{max}$); purple lines give the minimum cloud-edge raindrop radius required for the raindrop to not evaporate before reaching the surface ($r_\text{min}$) for three surface RH values. Purple shaded regions show cloud-edge size bounds of raindrops that can reach the surface. The gray shaded region sketches cloud condensation nuclei sizes.}
\label{fig:r_bounds_v_pcond}
\end{figure}

In Figure \ref{fig:r_bounds_v_pcond}, we calculated $r_\text{max}$ and $r_\text{min}$ for water raindrops across a range of planetary conditions. For clarity, we plot only one of the $r_\text{max}$ values from the methods we discussed previously (Rayleigh-Taylor with $\ell_\text{RT,max} = 0.5\pi a$) and $r_\text{min}$ values for only three representative surface relatives humidities (RH = 0.25, 0.5, 0.75) and no vertical wind ($w=0$). We used as default values the Earth-like conditions of Table \ref{tab:conditions} when the x-axis planetary parameter is not varying. 

The purple shading highlights how our estimates of $r_\text{max}$ and $r_\text{min}$ constrain possible raindrop sizes that can transport condensible mass from a cloud to the surface reservoir. Varying colors of purple shading correspond with size bounds set from varying surface relative humidity values. As the purple shading lightens, the bound includes smaller $r_\text{min}$ values associated with higher surface relative humidity values. The dark purple shading (for RH$_\text{surf}$=0.25) is the strictest bound.
For perspective, cloud drops grow from cloud condensation nuclei with many orders of magnitude smaller sizes as we have schematically indicated with gray shading. On Earth, typical cloud condensation nuclei are around 0.05 \textmu m ($5\times10^{-5}$ mm) \cite{Lohmann2016} while, for conditions considered here, viable raindrop sizes vary by about an order of magnitude with typical values of tenths of millimeters, or about 10,000 times larger than typical cloud condensation nuclei. 

In Figure \ref{fig:r_bounds_v_pcond}(a), we plot raindrop size bounds as described for variable dry surface pressures. Neither of the size bounds has a strong dependence on pressure (see also section \ref{subsec:r_max}). $r_\text{min}$ depends on pressure in multiple ways that largely cancel each other out. Figure \ref{fig:r_bounds_v_pcond}(b) shows the impact of surface gravity $g$ on raindrop size bounds. As $g$ increases, $r_\text{max}$ and $r_\text{min}$ all systematically decrease like $g^{-0.5}$. Larger raindrops are possible at lower surface gravities, and raindrops must also be larger to survive to the surface without evaporating. Figure \ref{fig:r_bounds_v_pcond}(c) highlights the effect of increased evaporation rate in higher air temperatures. 
$r_\text{min}$ rises with $T_\text{surf}$ because evaporation rate and falling time to surface increase as long as the molar mass of the condensible gas is less than the average dry air molar mass---as considered here. 

\subsection{Raindrop Evaporation with Varying Atmospheric Composition} 
Table \ref{tab:composition} considers the effect of atmospheric composition on the time and distance from cloud base until evaporation ($t_\text{evap}$ and $z_\text{evap}$, respectively) for \ce{H2O} raindrops with pure \ce{H2}, \ce{He}, \ce{N2}, \ce{O2}, and \ce{CO2} atmospheres. Other planetary conditions used are given under ``composition'' in Table \ref{tab:conditions}.
Atmospheric composition impacts raindrop evaporation from three main effects: (1) molar mass and heat capacity impact atmospheric structure, which governs how vertical distance maps to temperature, pressure, and relative humidity---all key parameters in calculating d$r$/d$z$; (2) molar mass impacts how a given pressure maps to a density, which impacts raindrop terminal velocity; and (3) molar mass and molecular structure impact the rate at which air can transport latent heat and condensible gas away from the raindrop. 

In Table \ref{tab:composition}, we calculated $t_\text{evap}$ and $z_\text{evap}$ considering each of these effects of composition in isolation as well as all together. When only a single effect is considered, all other compositional effects are calculated with pure \ce{N2}. Total number of molecules at cloud base is held fixed (i.e., pressure is fixed as an ideal gas assumed). 

As shown Table \ref{tab:composition}, air composition acts on the integral of d$r_\text{eq}$/d$z$ in competing directions, so all effects must be considered in unison to understand how precipitation evaporation will vary with atmospheric composition. 
We find the time taken to evaporate is comparable across atmospheric conditions. The distance to evaporation is comparable for the higher molecular mass gases with the \ce{He} atmosphere about 1.75 times larger and the \ce{H2} atmosphere about 3.5 times larger. Excluding the noble gas He, the ability to transport condensible mass in units of scale heights increases as atmospheric molar mass increases.

\begin{table}
\caption{\textit{Effects of Atmospheric Composition on Raindrop Evaporation}}
\begin{center}
\begin{tabular}{|l|l|l|l|l|}
\hline
composition & effect(s)                       & $t_\text{evap}$   [s] & $z_\text{evap}$   [m] & $z_\text{evap}$ [$H$] \\
\hline
\hline
\textbf{\ce{H2}}     & \textbf{all}                             & \textbf{769}                 & \textbf{6970}                & \textbf{0.0648}             \\
\hline
\ce{H2}     & $H$                             & 2610                & 7980              & 0.0741               \\
\hline
\ce{H2}     & $v_\text{T}$                           & 435                 & 3960                & 0.0368               \\
\hline
\ce{H2}     & transport & 333                 & 1000                & 0.00933               \\
\hline
\hline
\textbf{\ce{He}}     & \textbf{all}                             & \textbf{638}                 & \textbf{3560}                & \textbf{0.0632}               \\
\hline
\ce{He}     & $H$                             & 1570                & 4920               & 0.0820               \\
\hline
\ce{He}     & $v_\text{T}$                           & 623                   & 3480               & 0.0618               \\
\hline
\ce{He}     & transport & 329                 & 991               & 0.0176               \\
\hline
\hline
\textbf{\ce{N2}}     & \textbf{all}                             & \textbf{707}                 & \textbf{2090}              & \textbf{0.251}               \\
\hline
\hline
\textbf{\ce{O2}}     & \textbf{all}                            & \textbf{701}                 & \textbf{1900}               &  \textbf{0.260}               \\
\hline
\ce{O2}     & $H$                             & 675                 & 1920               & 0.263               \\
\hline
\ce{O2}     & $v_\text{T}$                           & 771                 & 2000               &  0.274               \\
\hline
\ce{O2}     & transport & 705                 & 2090               & 0.285               \\
\hline
\hline
\textbf{\ce{CO2}}    & \textbf{all}                             & \textbf{769}                 & \textbf{1960}               & \textbf{0.369}               \\
\hline
\ce{CO2}    & $H$                             & 655                 & 1870               & 0.351               \\
\hline
\ce{CO2}    & $v_\text{T}$                           & 726                 & 1750               & 0.330               \\
\hline
\ce{CO2}    & transport & 855                 & 2510               & 0.473\\
\hline
\end{tabular}
\caption*{\textit{Note.} Time to evaporate $t_\text{evap}$ and falling distance from cloud base before evaporation $z_\text{evap}$ in meters and relative to atmospheric scale height $H$ for a water raindrop of initial size {$r_0 = 0.5$} mm in different composition atmospheres. Planetary conditions besides dry composition are given in Table \ref{tab:conditions} under ``composition.'' With the effect column, we consider the three main impacts of composition on $t_\text{evap}$ and $z_\text{evap}$---atmospheric scale height $H$, raindrop terminal velocity $v_\text{T}$, and transport rate of condensible gas molecules and heat away from the raindrop surface ``transport''---together (``all'') as well as in isolation.}
\end{center}
\label{tab:composition}
\end{table}

\begin{figure}
\centering
\includegraphics[height=0.75\textheight]{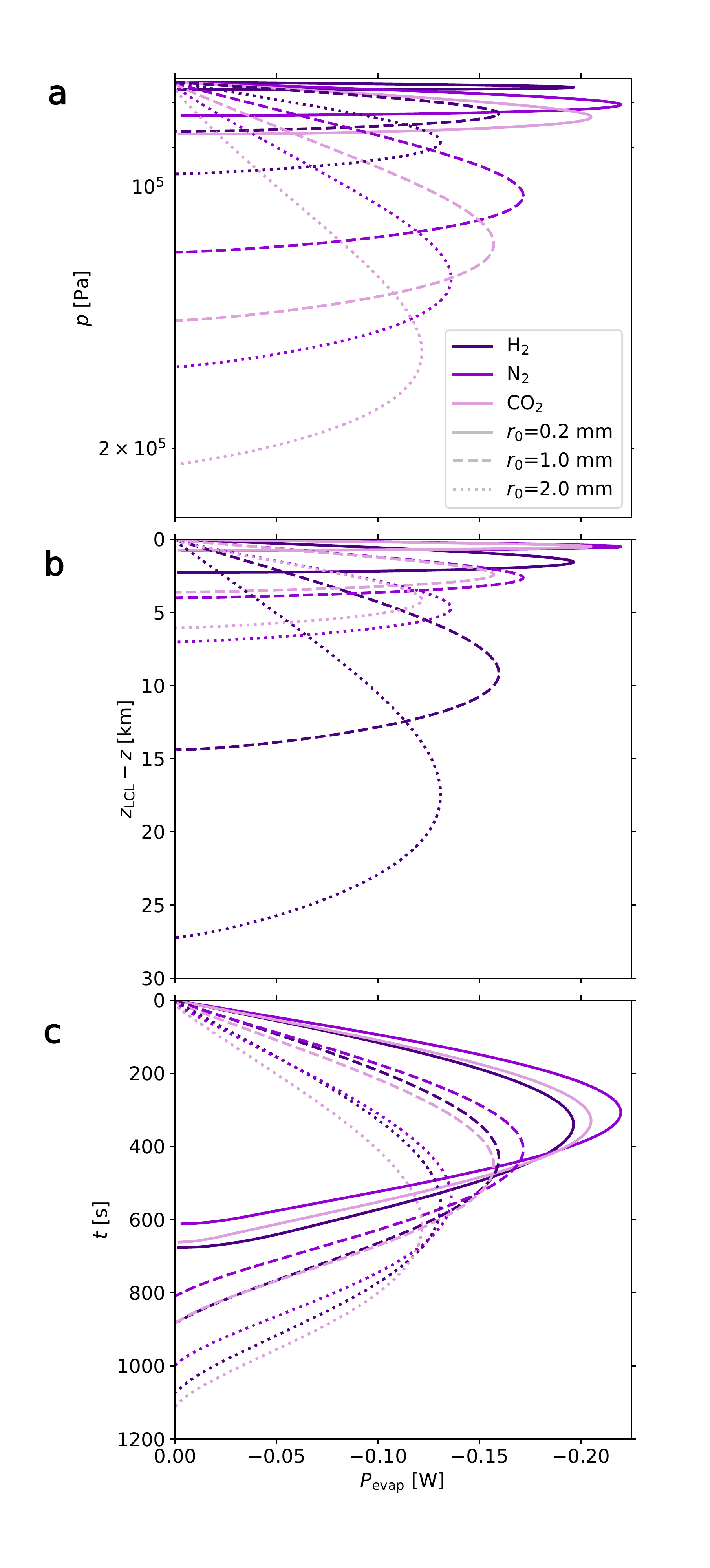}
\caption{(a) Pressure $p$ versus the rate of heat absorbed from local air $P_\text{evap}$ to evaporate the condensed water mass in a raindrop of size 2 mm distributed into raindrops of initial radii $r_0$ (varying line styles) in different atmospheric compositions (varying colors) falling from $T_\text{LCL}=275$ K and $p_\text{LCL} = 7.5\times10^{4}$ Pa under Earth surface gravity. (b) Same as (a) except falling distance from cloud base ($z_\text{LCL} - z$) versus $P_\text{evap}$. (c) Same as (a) except falling time $t$ versus $P_\text{evap}$.}
\label{fig:p_vs_P_evap}
\end{figure}

We further probe the variations in raindrop evaporation due to atmospheric composition in Figure \ref{fig:p_vs_P_evap} by calculating latent heat absorbed per second (power, $P_\text{evap}$) as a function of (a) atmospheric pressure, (b) vertical distance from cloud base, and (c) falling time for a fixed initial condensible mass sorted into raindrops of three different radii. $P_\text{evap}$ is linearly related to the rate of condensible mass evaporation through the latent heat of vaporization, which is normalized per unit mass. Thus we choose a constant initial mass (the mass of the largest raindrop considered) to compare magnitudes of $P_\text{evap}$ for different initial raindrop sizes. We use the same planetary conditions as Table \ref{tab:composition} (under ``composition'' in Table \ref{tab:conditions}) with background \ce{H2}, \ce{N2}, and \ce{CO2} atmospheres.

In all three vertical coordinates shown in Figure \ref{fig:p_vs_P_evap}, the peak $P_\text{evap}$ reached is about the same for a given $r_0$ across all three atmospheric compositions. Maximum $P_\text{evap}$ decreases as $r_0$ increases (as expected in order to conserve mass with longer fall times). Figure \ref{fig:p_vs_P_evap} shows $P_\text{evap}$ is roughly a quadratic function of log $p$ (a), $z_\text{LCL}-z$ (b), and $t$ (c) between cloud base and reaching total evaporation for all atmospheric compositions and initial size values considered. 

For a given composition atmosphere, increasing $r_0$ increases $p_\text{evap}$, $z_\text{evap}$, and $t_\text{evap}$. The spread in $p_\text{evap}$ and $z_\text{evap}$ at total evaporation between different compositions increases as $r_0$ increases. As seen in Table \ref{tab:composition} (given log $p$ is essentially proportional to altitude over scale height), $p_\text{evap}$ increases with increasing dry molar mass; $z_\text{evap}$ decreases with increasing dry molar mass; and $t_\text{evap}$ is about constant across the different compositions.

\subsection{Comparison of Water Raindrops on Specific Planets}
\begin{figure}
\centering
\includegraphics[height=0.75\textheight]{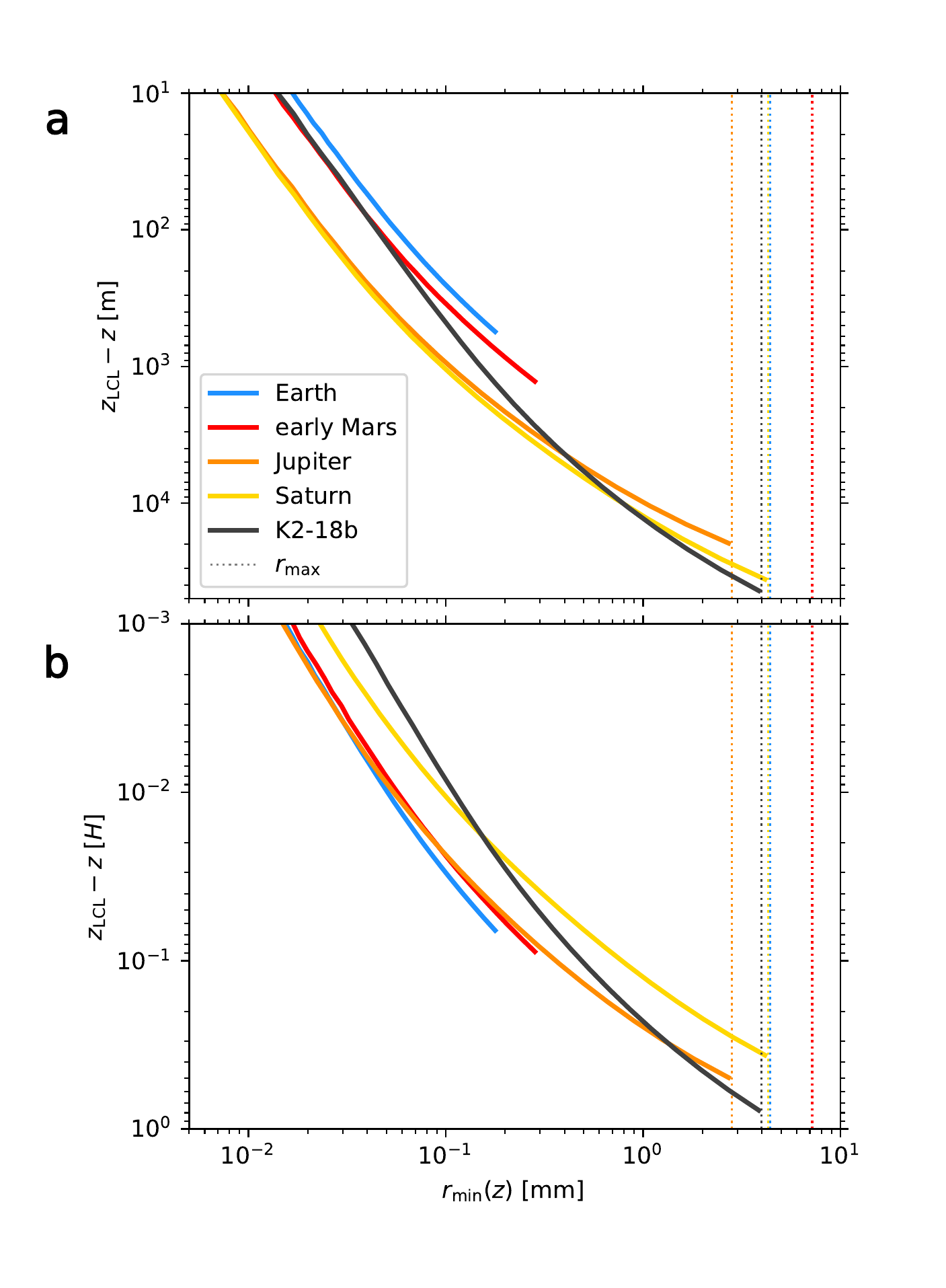}
\caption{Distance from cloud base $z_\text{LCL} - z$ in units of (a) meters and (b) atmospheric scale heights versus the minimum initial radius raindrop to reach altitude $z$ without total evaporation ($r_\text{min}(z)$). Different colored solid lines represent ostensible atmospheric conditions (given by Table \ref{tab:conditions}) for different planets as labeled. All raindrops are composed of water. Thin dashed vertical lines indicate maximum stable raindrop radius $r_\text{max}$ as estimated via Rayleigh-Taylor instability with $\ell_\text{RT,max}=0.5\pi a$, distinguished by planet with labeled colors. $r_\text{min}(z)$ values are plotted until $r_\text{min}(z) = r_\text{max}$ or $z$ intersects with the planet's surface.}
\label{fig:spec_rmin}
\end{figure}
Next we move from considering raindrops in abstract conditions to studying raindrops in known (or speculated) planetary conditions.
In Figure \ref{fig:spec_rmin}, we compare water raindrops on Earth; warm, wet ancient Mars; Jupiter; Saturn; and exoplanet K2-18b. Warm, wet ancient Mars is a hypothesized climate state 3-4 billion years ago where Mars was warm compared to the melting point of water and rainfall was frequent \cite<e.g.,>[]{Wordsworth2016}.

K2-18b is an exoplanet without analog in the solar system, falling between the sizes of Earth and Neptune and receiving an Earth-like amount of stellar insolation \cite{Foreman2015,Montet2015,Cloutier2019,Benneke2019}. Though many of K2-18b's characteristics are imprecisely known, we selected this exoplanet to demonstrate the flexibility of our model as multiple teams have claimed observational detections of water vapor \cite{Tsiaras2019,Benneke2019,Madhusudhan2020} and one team has hypothesized that observations suggest the presence of liquid water clouds \cite{Benneke2019}.

We set surface conditions on the terrestrial Earth and Mars and cloud base conditions on the gaseous Jupiter, Saturn, and K2-18b as given in Table \ref{tab:conditions}. 
We plot altitude $z$ from cloud base in units of (a) meters and (b) atmospheric scale heights versus $r_\text{min}(z)$, the minimum radius raindrop to reach $z$ without totally evaporating. How $r_\text{min}(z)$ values cluster among planets varies depending on whether the distance from cloud base is measured in scale heights or meters, but for all $z$ values tested in both unit systems we find $r_\text{min}(z)$ varies among the diverse planetary conditions considered by only a factor of few. This agreement has interesting implications for future work aimed at rigorous generalization of raindrop microphysics schemes from Earth to other planets.

\subsection{Raindrops beyond \ce{H2O}}
While water is the most familiar liquid condensible, outside of the familiar Earth temperature range, many other species condense as liquids and can form raindrops. Table \ref{tab:raindrop_comp} compiles a number of liquid condensible species that are cosmochemically abundant. Detailed analysis of raindrop behavior requires specifying atmospheric conditions in addition to condensible species, so here we consider how basic condensible properties vary raindrop behavior relative to water. Again in the interest of simplicity, we do not consider mixtures of condensibles predicted by thermodynamic equilibrium in many atmospheric gas combinations, e.g., \ce{N2}-\ce{CH4} or \ce{NH3}-\ce{H2O} \cite{Thompson1992,Guillot2020a}.

We give the melting temperature $T_\text{melt}$ for each condensible to give an idea of the atmospheric temperatures where each species will be liquid. Lower temperature condensibles like \ce{CH4} and \ce{NH3} dominate the observable clouds of the outer solar system. The \ce{CH4} cycle of Titan is the only active ``terrestrial'' condensible cycle besides Earth's we can observe in detail. 

Metal and rock species like \ce{Fe} and \ce{SiO2} become condensible species at very high temperatures. Such species are predicted to be condensibles on highly irradiated exoplanets that are favored observational targets for the foreseeable future; \citeA{Ehrenreich2020} have already claimed observational evidence of \ce{Fe} condensing on WASP-76b. On Earth, such high temperatures can be reached during asteroid/meteoroid impacts. Geologically preserved impact spherules \cite<e.g.,>[]{Johnson2012a,Johnson2012b} and micrometeorites \cite<e.g.,>[]{Tomkins2016,Payne2020} both undergo phases in their life through the atmosphere where their behaviors are described by the raindrop physics we have presented. 

Despite the very wide range of condensible species, $r_\text{max}$ values in Table \ref{tab:raindrop_comp} only vary relative to water by a factor of 0.5 to 2. 
As we have examined in section \ref{subsec:r_max}, the only planetary parameter beyond condensible type that affects $r_\text{max}$ is surface gravity $g$, which for planetary bodies varies about an order of magnitude (between about 1 and 25 m s$^{-1}$). We therefore find that maximum stable raindrop sizes are remarkably similar across a very wide range of planetary conditions and raindrop compositions.

\begin{table}[]
\caption{\textit{Properties of Liquid Condensibles and Their Raindrops}}
\begin{tabular}{|l|l|l|l|l|l|l|}
\hline
condensible & $T_\text{melt}$ & $\rho_\ell$  & $\sigma$  & $r_\text{max}$ & $L$ & $E_\text{evap}(r)$  \\
& [K] & [kg m$^{-3}$] & [N m$^{-1}$] & $[r_{\text{max}}(\ce{H2O})]$ & [MJ    kg$^{-1}$] &  [$E_\text{evap}(r,\ce{H2O})$]\\
\hline
\ce{CH4}    & 91$^{a}$                  & 451$^{a}$                       & 0.0187$^{b}$                & 0.742                                     & 0.531$^{a}$                    & 0.0958                                             \\
\hline
\ce{NH3}    & 194$^{a}$                 & 733$^{a}$                       & 0.0445$^{b}$                & 0.897                                     &  1.49$^{c}$ & 0.437                                              \\
\hline
\ce{H2O}    & 273$^{a}$                 & 1000$^{a}$                      & 0.0754$^{d}$                 & 1                                         & 2.50$^{c}$                     & 1                                                  \\
\hline
\ce{Fe}     & 1811$^{e}$                & 7030$^{e}$                      & 1.92$^{f}$                  & 1.90                                     & 6.76$^{g}$                      & 19.0                                               \\
\hline
\ce{SiO2}   & 1996$^{h}$               & 2140$^{i}$                      & 0.3$^{j}$                   & 1.36                                      & 12.4$^{h}$                     & 10.6                                               \\
\hline 
\end{tabular}
\caption*{\textit{Note.} Temperature-dependent values use {$T=T_\text{melt}$}. $^{a}$\citeA{NIST}. $^{b}$\citeA{Somayajulu1988}. $^{c}$\citeA{Rumble2017}. $^{d}$\citeA{Vargaftik1983}. $^{e}$\citeA{Assael2006}. $^{f}$\citeA{Brillo2005}. $^{g}$\citeA{Desai1986}. $^{h}$\citeA{Melosh2007}. $^{i}$\citeA{Bacon1960}. $^{j}$\citeA{Kingery1959}.}
\label{tab:raindrop_comp}
\end{table}

\section{Discussion}\label{sec:diss} 
This work is merely a first step toward a generalized theory of how precipitation and condensible cycles operate in planetary conditions different from modern Earth. We have considered only single raindrops, independent of their formations. Future progress will require development of theory for general planetary atmospheres on the growth of raindrops from cloud drops and extensions that include solid precipitating particles and their growth. 
In this context, below we discuss some future applications and extensions of this work.

\subsection{Precipitation Efficiency}\label{subsec:precip_eff}
Precipitation efficiency measures how efficiently an atmosphere transports condensed mass from a cloud downward. Qualitatively, its distribution over time is an important metric for planetary climate as it shapes cloud coverage (both temporally and spatially), cloud radiative properties, and relative humidity profiles, which all have large consequences for radiative balance \cite{Romps2014,Zhao2016,Lutsko2018}. On a terrestrial planet, precipitation efficiency evaluated at the surface is a particularly important quantity as it helps to set the amount of condensible mass in the atmosphere. 

On short timescales, liquid precipitation efficiency is governed by the basic raindrop physics of falling and evaporation we have presented here. We have demonstrated that the dimensionless number $\Lambda$ captures the essential behavior of raindrop evaporation and descent. For $\Lambda$ evaluated with the length-scale from cloud base to the surface, raindrops of initial sizes with $\Lambda>1$ (i.e., $r_0<r_\text{min}$) will totally evaporate and have no condensible mass reach the surface; raindrops of initial sizes with $\Lambda<0.1$ will experience little evaporation and have the majority of their mass reach the surface; and raindrops of initial sizes with $0.1<\Lambda\le1$ will have both condensible mass evaporate and reach the surface. Short-scale precipitation efficiency is then fundamentally controlled by the cloud-edge condensed mass distribution among these three different size categories.

How precipitation efficiency on short timescales---governed directly by microphysics---maps to the climatically important temporal distribution of precipitation efficiency is fundamentally influenced by both large-scale atmospheric dynamics and local-scale convection \cite{Romps2014}. Predicting precipitation efficiency distributions in three-dimensional models requires capturing key microphysical behaviors in a parameterized representation. Our analysis suggests the key microphysical behavior we must capture to physically represent precipitation efficiency on short timescales is the sorting of condensed cloud mass into the three different size categories determined by $\Lambda$.  
This interpretation suggests a physical grounding for microphysics' role in controlling an important yet poorly understood climate parameter \cite{Zhao2016,Lutsko2018}. Tying precipitation efficiency to the raindrop size distribution may therefore provide a framework for future improvements in generalized microphysics parameterizations.

Given this analysis, the bounds on water raindrop sizes in Figure \ref{fig:r_bounds_v_pcond} show one component of how precipitation efficiency will be shaped by different planetary conditions. A complete picture of precipitation efficiency obviously will also require better understanding of how planetary parameters influence the formation and subsequent mass distribution of raindrops.
Nevertheless, the shrinking viable surface-reaching raindrop size range with rising $T_\text{surf}$ in Figure \ref{fig:r_bounds_v_pcond}(c) is a striking predicted feature of warmer water cycles, which may have implications for a \ce{CO2}-rich early Earth, for early Venus, or for exoplanets close to the runaway greenhouse threshold.

\subsection{Convective Storm Dynamics}
Evaporating raindrops also influence convective storm dynamics. The vertical transport of heat and condensible mass from evaporating raindrops causes local variations in air density through changes in both average molar mass and temperature. The implications of evaporating raindrops for convection depend on the ratio between an atmosphere's dry mean molecular mass $\mu_\text{dry}$ and the molecular mass of its condensing species $\mu_\text{c}$ as well as the removal rate of latent heat relative to the local $T-p$ profile \cite{Guillot1995,Leconte2017}. 

On modern-Earth, $\mu_\text{c}/\mu_\text{dry} \approx 0.6$, so molar-mass-contrast effects are present but muted. Nonetheless, the interplay of rising air supplying condensible mass and sinking precipitating air plays a key role in the evolution and lifetime of a given storm \cite<e.g.,>[]{Rogers1996}. Variations in $\mu_\text{c}/\mu_\text{dry}$ have been hypothesized to drive storm systems completely unlike those on Earth---e.g., Saturn's giant white storms \cite{Li2015,Leconte2017}. 

As suggested by Table \ref{tab:composition} and Figure \ref{fig:p_vs_P_evap}, an atmosphere's background dry gas also influences the ability of raindrops to vertically re-distribute latent heat and condensible mass with respect to $z$ and log $p$. This effect of background gas composition will also likely influence air density changes during storms. Future dynamical studies might explore the implications of these variations in raindrop evaporation with background gas for storm evolution and subsequent condensible gas distributions. One example solar system application of interest is better constraining how deep and how effectively ammonia raindrops (originating from melted \ce{NH3} snowflakes or hail-like \ce{NH3-H2O} ``mushballs'') can transport ammonia on Jupiter \cite{Ingersoll2017,Li2019,Li2020,Guillot2020a,Guillot2020b}.

\subsection{Rainfall Rates}
On terrestrial planets, rainfall rate---the mass flux of liquid condensible hitting the surface---is a key characteristic for predicting surface erosion and flooding events \cite<e.g.,>[]{Kavanagh2015,Craddock2017,Margulis2017}. Rainfall rate depends on liquid condensible mass per unit air, the raindrop size distribution, and raindrop velocities as a function of size \cite{Rogers1996}. 
While we cannot yet make robust predictions for how rainfall rates should vary in different planetary conditions, we expect such a relationship will be sensitive to $g$ and air density at cloud level because of the role of collisional kinetic energy in shaping raindrop size distributions \cite{Low1982,Rogers1996,Pinsky2001,List2009}. $r_\text{min}$ and $\Lambda$ as a function of $r_\text{eq}$ evaluated at the surface are useful for contextualizing the size range of the in-cloud raindrop size distribution that is important for predicting rainfall rates. Future work should investigate how narrow bounds of viable surface-reaching raindrops might constrain rainfall rates across different planetary conditions in more detail. 

\subsection{Extension to Solid Particles}
In this work, we have focused on falling liquid condensible particles because solid particle shapes are much more complicated and variable than liquid oblate spheroids. 
The morphology of solid condensed particles exhibits extreme variability because of the high sensitivity of crystal orientation to temperature and condensible vapor super-saturations \cite<e.g.,>[]{Libbrecht2017}. Further, crystal structure is fixed at deposition, so because a solid condensed particle experiences variations in environmental conditions during its growth, its final shape is highly sensitive to its growth path. These considerations mean generalizing modern-Earth modeling approaches for handling ice shape degeneracies \cite<e.g.,>[]{Krueger1995} is a non-trivial exercise. 

Understanding of shape is the main limiting factor for applying the methodology presented here for raindrops to solid condensible particles. Shape is a key parameter in calculating $C_D$ and $f_V$, which impact terminal velocity and evaporation rate, respectively. Shape also needs to be accounted for more fundamentally in the derivation of evaporation rate because it controls boundary conditions, but analogous mathematics to well-investigated electrostatics means such boundary condition accounting has already been compiled \cite<e.g.,>[chapter 8]{Mcdonald1963,Lamb2011}. Extension to solid particles would also need to account for differences in saturation pressure with respect to solid and liquid condensibles and a larger assortment of latent heats because of more available phase changes, but these issues are a question of compiling thermodynamic data rather than an inherent lack of understanding.  

\section{Conclusion}\label{sec:con}
We have compiled and generalized methods for calculating raindrop shape, terminal velocity, and evaporation rate in any planetary atmosphere. These properties govern raindrop behavior below a cloud, a simple but necessary component of understanding how condensible cycles operate across a wide range of planetary parameters beyond modern-Earth conditions. For terrestrial planets, raindrops sizes capable of transporting condensed mass to the surface only span about an order of magnitude, a narrow bound when compared to origins in cloud condensation nuclei many orders of magnitude smaller. We show across a wide range of condensibles and planetary parameters that maximum raindrop sizes do not significantly vary. 

With more in depth calculations, we confirm the conclusion of \citeA{Palumbo2020} that maximum raindrop size is only weakly dependent on air density, in contrast to the results of an earlier study \cite{Craddock2017}. By returning to the physics equations governing raindrop falling and evaporation, we demonstrate raindrop ability to vertically transport latent heat and condensible mass can be well captured by a new dimensionless number. Our analysis suggests cloud-edge, mass-weighted raindrop size distribution is a key microphysics-based control on the important climate parameter of precipitation efficiency. 

\acknowledgments

Our model and all code used to generate results for this paper are available at an archived Github repository \cite<>[\url{https://github.com/kaitlyn-loftus/rainprops}]{Loftus2021}. 
This work was supported by NSF grant AST-1847120. K.L. thanks Mikl\'{o}s Szak\'{a}ll for providing data on raindrop shape experiments as well as Mark Baum, Ralph Lorenz, and Jacob Seeley for helpful discussions pertaining to numerical methods, Martian raindrops, and climatic impacts of microphysics parameterizations, respectively. We thank Tristan Guillot and Ralph Lorenz for insightful reviews.

\begin{notation}
\notation{$A$} [m$^2$] cross sectional area
\notation{$a$} [m] semi-major axis of oblate spheroid
\notation{$b$} [m] semi-minor axis of oblate spheroid
\notation{$C_\text{D}$} [ ] drag coefficient
\notation{$C_\text{shape}$} [ ] oblate spheroid shape correction in drag coefficient
\notation{$c_p$} [J K$^{-1}$ kg$^{-1}$] specific heat capacity at constant pressure 
\notation{$D_\text{c-air}$} [m$^2$ s$^{-1}$] diffusion coefficient of condensible gas in air  
\notation{$E$} [J] energy 
\notation{$F_\text{x}$} [N] force specified by subscript
\notation{$f_\text{SA}$} [ ] ratio of the surface area of an oblate spheroid to a sphere of equivalent radius 
\notation{$f_\text{V}$} [ ] ventilation coefficient, subscript can further specify for molecular or heat transport 
\notation{$g$} [m s$^{-2}$] planetary gravity
\notation{$H$} [m] atmospheric scale height
\notation{$K_\text{air}$} [W m$^{-1}$ K$^{-1}$] thermal conductivity of air
\notation{$L$} [J kg$^{-1}$] latent heat of vaporization
\notation{$\ell$} [m] length scale
\notation{$m$} [kg] mass of raindrop

\notation{$P$} [W] power
\notation{$p$} [Pa] air pressure
\notation{$p_\text{sat}$} [Pa] saturation pressure of condensible gas
\notation{Pr} [ ] Prandtl number 
\notation{$R$} [J mol$^{-1}$ K$^{-1}$] ideal gas constant
\notation{Re} [ ] Reynolds number
\notation{RH} [ ] relative humidity
\notation{$r_0$} [m] initial raindrop radius at cloud base
\notation{$r_\text{eq}$} [m] equivalent radius
\notation{$r_\text{max}$} [m] maximum raindrop radius before breakup
\notation{$r_\text{min}$} [m] minimum threshold raindrop radius to reach without total evaporation a given distance from cloud base; if no length scale is specified, assumed to be distance to surface
\notation{Sc} [ ] Schmidt number
\notation{$T$} [K] temperature
\notation{$t$} [s] time
\notation{$v_\text{T}$} [m s$^{-1}$] terminal velocity
\notation{$w$} [m s$^{-1}$] vertical wind speed
\notation{$z$} [m] vertical space coordinate

\notation{$\beta$} [ ] exponential dependence of raindrop velocity on raindrop radius 
\notation{$\Delta r$} [m] threshold radius size for drop to be considered ``fully evaporated'' 
\notation{$\Delta T_\text{drop}$} [K] steady-state temperature difference between evaporating raindrop and local air 
\notation{$\eta$} [Pa s] dynamic viscosity
\notation{$\Lambda$} [ ] dimensionless number describing raindrop evaporative mass loss
\notation{$\lambda$} [m] wavelength
\notation{$\mu$} [kg mol$^{-1}$] molar mass
\notation{$\sigma$} [N m$^{-1}$] surface tension
\notation{$\rho$} [kg m$^{-3}$] density

\notation{$X_\text{air}$} describes air property at local altitude
\notation{$X_\text{c}$} describes a condensible property 
\notation{$X_\text{drop}$} describes a raindrop property 
\notation{$X_\ell$} describes a liquid condensible property where ambiguous if ``c'' references liquid or gas condensible
\notation{$X_\text{evap}$} describes property when raindrop finishes evaporating
\notation{$X_\text{LCL}$} describes air property at lifting condensation level
\notation{$X_\text{surf}$} describes air property at surface
\end{notation}

\appendix

\section{Oblate Spheroid Geometrical Relationships}\label{app:oblate}
Here, we give geometric properties of our assumed oblate-spheroid raindrop with semi-major axis $a$, semi-minor axis $b$, and equivalent radius $r_\text{eq}$. Note that a sphere is an oblate spheroid of $b/a=1$.

Raindrop semi-major axis $a$ can be determined numerically for a given $r_\text{eq}$ and $b/a$ from the relationship
\begin{linenomath*}
\begin{equation}\label{eq:a_f_r_eq}
    a = r_\text{eq}\left(\frac{b}{a}\right)^{-1/3}
\end{equation}
\end{linenomath*}
\cite{Green1975}. An oblate spheroid has cross sectional area $A$ of
\begin{linenomath*}
\begin{equation}
    A = \pi a^2 = \pi r_\text{eq}^2 \left ( \frac{b}{a}\right )^{-\frac{2}{3}}
\end{equation}
\end{linenomath*}
\cite{Green1975} and a volume $V$ of
\begin{linenomath*}
\begin{equation}
    V = \frac{4}{3}\pi a^2 b = \frac{4}{3}\pi r_\text{eq}^3,
\end{equation}
\end{linenomath*}
by definition of $r_\text{eq}$.
The ratio of the surface area of an oblate spheroid to that of a sphere $f_\text{SA}$ is
\begin{linenomath*}
\begin{equation}
f_\text{SA} = 
\begin{cases}
   0.5\left(\frac{b}{a}\right)^{-2/3} + \left(\frac{b}{a}\right)^{4/3}\left(4\epsilon\right)^{-1}\ln\left[\frac{1+\epsilon}{1-\epsilon}\right], & b/a < 1\\
   1, & b/a = 1\\ 
\end{cases}
\end{equation} 
\end{linenomath*}
where $\epsilon = \sqrt{1-(b/a)^2}$ \cite{Loth2008}.

\section{Raindrop Shape}\label{app:shape}
The shape of a falling raindrop in air (i.e., the surface of the condensible-air boundary) can be described by the Young-Laplace equation, which governs the surface boundary between two immiscible (non-mixing) fluids. For a given point on a raindrop surface, the Young-Laplace equation can be written as 
\begin{linenomath*}
\begin{equation}\label{eq:Y-L}
  \sigma_{\text{c-air}} \left( R_1^{-1} + R_2^{-1} \right) = \Delta p 
\end{equation}
\end{linenomath*}
\cite<see>[chapter 10.3.2 for derivation]{Pruppacher2010}. 
$\sigma_{\text{c-air}}$ is surface tension between liquid condensible and air; $R_1$ and $R_2$ are the principle radii of curvature, which together describe the shape of the local surface via its curvature \cite{deGennes2013}; and $\Delta p$ is the difference between internal and external pressures on either side of the raindrop's surface boundary.

Because we have pre-assumed an oblate spheroid geometry, we only have to evaluate equation \eqref{eq:Y-L} at one point on the raindrop's surface rather than integrating over the entire surface, and the principle radii are analytic expressions rather than values which must be iteratively corrected for such integration to be self consistent \cite<see>[for shape calculation without oblate spheroid assumption]{Beard1987}. We follow \citeA{Green1975} and solve equation \eqref{eq:Y-L} at a point on the raindrop's equator. From geometry, for all points on an oblate spheroid's equator, 
\begin{linenomath*}
\begin{equation}
    R_1 = b^2 a^{-1}, \label{eq:R1}
\end{equation}
\end{linenomath*}
and
\begin{linenomath*}
\begin{equation}
     R_2 = a \label{eq:R2}
\end{equation}
\end{linenomath*}
\cite<see>[Appendix B for derivation]{Green1975}.

In a full accounting of pressures, internal pressure $p_\text{int}$ should include hydrostatic pressure, pressure from spherical surface tension, and pressure from internal circulation within the raindrop. External pressure $p_\text{ext}$ should include hydrostatic pressure, pressure from aerodynamic drag, and pressure from air turbulence. Following \citeA{Green1975}, we only consider internal and external hydrostatic pressures and pressure from spherical surface tension, giving
\begin{linenomath*}
\begin{equation}\label{eq:delta_p_full}
    \Delta p = p_\text{int} - p_\text{ext} \approx (\rho_{\text{c}}-\rho_\text{air})gb + 2\sigma_{\text{c-air}}r_\text{eq}^{-1}
\end{equation} 
\end{linenomath*}

Pressure from drag is fundamental to shaping the raindrop's shape at larger raindrop sizes (i.e., where the magnitude of $F_\text{drag}$ begins to approach the magnitude of $F_\sigma$); drag is responsible for the evolution in raindrop shape from oblate spheroid to upper hamburger bun. Drag acts to deform the raindrop so that it is no longer axially symmetric about the major axis because drag is not uniformly distributed over the raindrop's surface. 

Our assumption of oblate spheroid shape is incompatible with a detailed consideration of the effects of drag on raindrop shape. We fully neglect drag because we consider the pressure balance at raindrop equator where drag pressure is minimal even when the total magnitude of the drag force on the raindrop is significant \cite{Green1975}. The consequences of this neglect can be seen to be minimal from comparison of \citeA{Green1975}'s oblate spheroid method to computed or observed raindrops shapes where drag is included \cite{Beard1987,Thurai2009}.

We neglect the effects of atmospheric turbulence on raindrop shape as we are calculating an equilibrium shape; in practice, turbulence acts to induce oscillations in raindrop shape (which are then viscously damped) rather than changing the equilibrium shape \cite{Beard2010}.
We also neglect the effects of internal circulation within the raindrop as empirically its neglect has no impact on predicting Earth raindrop shapes \cite{Thurai2009} and theoretically internal circulation is expected to be small for a higher dynamic viscosity liquid raindrop falling through a lower dynamic viscosity air \cite{Clift2005}. %pg 125 

\section{Maximum Raindrop Size before Breakup}\label{app:breakup}

\subsection{Rayleigh-Taylor instability}
Raindrop breakup can be studied by a linear instability analysis that incorporates capillary and gravity waves. 
For wavelengths above a critical wavelength $\lambda^\ast$, total wave phase velocity on the base of the drop becomes imaginary; waves rapidly amplify in magnitude; and the raindrop becomes unstable. Assuming planar surface waves (the 3D corrections for drops near the size of $r_\text{max}$ are small \cite{Dhir1973,Grace1978}), the critical wavelength can be written as 
\begin{linenomath*}
\begin{equation}
    \lambda^\ast = 2 \pi \sqrt{\frac{\sigma_{\text{c-air}}}{g(\rho_{\text{c},\ell}- \rho_\text{air})}}
\end{equation}
\end{linenomath*}
\cite{Komabayasi1964,Grace1978}.

Physically, for a raindrop to not be disrupted by a wave, the zero mode of wavelength $\lambda^\ast/2$ must be less than the characteristic size of the drop:
\begin{linenomath*}
\begin{equation}\label{eq:ell_max_rt_app}
    \ell_\text{RT,max} = \frac{\lambda^\ast}{2} = \pi \sqrt{\frac{\sigma_{\text{c-air}}}{g(\rho_{\text{c},\ell} - \rho_\text{air})}}.
\end{equation}
\end{linenomath*}
Different metrics exist for the maximum raindrop length scale relative to the zero mode wave $\ell_\text{RT,max}$---e.g., \citeA{Grace1978} uses half the equivalent radius' circumference, $0.5\pi r_\text{eq}$, while \citeA{Pruppacher2010} uses the maximum physical raindrop diameter, $2a$. 
Different choices of length scale will result in quantitatively different results for the onset of instability by a factor of a few; this discrepancy emphasizes the role of this analysis as an \textit{estimate} of when drops tend to become unstable. Regardless, any choice of length scale results in the same dependencies on physical parameters---our primary concern here. Length scales using semi-major axis $a$ can be related to $r_\text{eq}$ via oblate spheroid geometry.

While no experimental pressure chamber data exists to explicitly test air density's effect on maximum raindrop size,
numerous chemical engineering experiments have been done on the maximum-sized drops of different media falling through various other media. 
Such experiments consistently agree (within about 20\%) with the predictions of equation \eqref{eq:ell_max_rt} where ``c'' is replaced with the drop medium, ``air'' is replaced with the fall medium, and the dynamic viscosity of the drop medium is much larger than the fall medium's (like raindrops and air) \cite{Lehrer1975,Grace1978,Clift2005}.
More complex wave analysis does not consistently produce better agreement with experiments \cite{Grace1978}.

\subsection{Surface Tension-Drag Force Balance}
The force on a raindrop due to surface tension is 
\begin{linenomath*}
\begin{equation}\label{eq:f_surface_t}
    F_\sigma = \ell_\sigma\sigma_{\text{c-air}}
\end{equation}
\end{linenomath*}
where $\ell_\sigma$ is a characteristic length scale of the raindrop's surface, which is conventionally taken to be $2\pi a$ or $2 \pi r_\text{eq}$. Again, there is ambiguity in length scales because this setup is an estimate rather than a rigorous calculation. Drag force is given by equation \eqref{eq:f_drag}.
We note the maximum velocity relative to air of a raindrop falling in isolation is its terminal velocity (where $F_\text{drag} = F_g$), and raindrop velocity is a monotonically increasing function with raindrop size. Thus, at the smallest $r_\text{eq}$ where $F_\text{drag} = F_\sigma$, the setup is analogous to solving for the raindrop size where the force of surface tension equals the gravitational force, given by equation \eqref{eq:f_g}.

\section{Analysis with $\Lambda$ of the Dependence of Raindrop Evaporation on Size}\label{app:nond}
We can understand the behavior of raindrop evaporation with changing raindrop size by considering the dependence of $\Lambda$ on $r_\text{eq}$. From equation \eqref{eq:Lambda_expanded}, we can approximately represent $\Lambda$ explicitly in terms of $r_\text{eq}$ as
\begin{linenomath*}
\begin{equation}\label{eq:simp_Lambda}
    \Lambda \approx C r_\text{eq}^{-(2+\beta)}.
\end{equation}
\end{linenomath*}
$C$ is a proportionality constant dependent on the environment across $\ell$ and the species of condensible; $\beta$ is defined such that d$z$/d$t \propto r_\text{eq}^\beta$. (We have neglected the dependence of ventilation factors on $r_\text{eq}$ to get a tractable expression here.)

For simplicity, we consider the limiting case where $w=0$, so d$z$/d$t=v_T$. For very small raindrops (Re$\ll$1), the raindrop is in the Stokes regime and $\beta=2$. For very large raindrops (as $r_\text{eq}$ approaches $r_\text{max}$), raindrop terminal velocity approaches a constant value \cite<>[chapter 7.C]{Clift2005}, and $\beta$ approaches 0. From the behavior of $C_D$, $\beta$ smoothly varies so that $\beta \in [0,2]$ \cite<e.g.,>[chapter 7.2.3]{Lohmann2016}. Therefore, as $r_\text{eq}$ decreases, $\Lambda$ always exponentially increases, with the dependence on $r_\text{eq}$ moving from $\Lambda \propto r_\text{eq}^{-2}$ to $\Lambda \propto r_\text{eq}^{-4}$. The exponential dependence of $\Lambda$ on $r_\text{eq}$ means that the transition from minimal evaporation to full evaporation occurs over a narrow raindrop size range.

One perspective on the width of this transition regime---applicable for raindrop sizes varying by orders of magnitude---is to consider the ratio of the $r_\text{eq}$ such that $\Lambda=0.1$ to the $r_\text{eq}$ such that $\Lambda=1$ (i.e., the ratio of sizes between the drop that evaporates 10\% of its initial mass and the drop that fully evaporates just as it reaches the end of the prescribed length scale). From equation \eqref{eq:simp_Lambda} and assuming $\beta$ is constant between $\Lambda=1$ and $\Lambda=0.1$, this ratio is equal to $10^{1/(2+\beta)}$, which evaluates to 1.78 and 3.16 for $\beta=2$ and $\beta=0$, respectively. Because $\beta$ decreases with increasing $r_\text{eq}$, as the $r_\text{eq}$ that evaluates to $\Lambda=1$ increases (e.g., by increasing $\ell$), the transition width increases. Still, regardless of the exact value of $\beta$ within [0,2], a change in raindrop size of a factor of 2-3 is small compared raindrop growth processes that require many order of magnitude changes in drop size.

Similar analysis featuring $\Lambda$ can be used to introduce the additional complexity of non-zero vertical wind speed or to isolate other variables of interest beyond $r_\text{eq}$ and consider their effects on raindrop evaporation.

\bibliography{bib}

\end{document}

%% ------------------------------------------------------------------------ %%
%
%  TITLE
%
%% ------------------------------------------------------------------------ %%

\title{Supporting Information for ``The Physics of Falling Raindrops in Diverse Planetary Atmospheres''}

DOI: 10.1002/TBD%insert paper number here%

%% ------------------------------------------------------------------------ %%
%
%  AUTHORS AND AFFILIATIONS
%
%% ------------------------------------------------------------------------ %%

\authors{=K. Loftus\affil{1} and R.D. Wordsworth\affil{1,2}=}

\affiliation{1}{Department of Earth and Planetary Sciences, Harvard University, Cambridge, MA, US}
\affiliation{2}{School of Engineering and Applied Sciences, Harvard, University, Cambridge, MA, US}

%% ------------------------------------------------------------------------ %%
%
%  BEGIN ARTICLE
%
%% ------------------------------------------------------------------------ %%

% The body of the article must start with a \begin{article} command
%
% \end{article} must follow the references section, before the figures
%  and tables.

\begin{article}

%% ------------------------------------------------------------------------ %%
%
%  TEXT
%
%% ------------------------------------------------------------------------ %%

\noindent\textbf{Contents of this file}
\begin{enumerate}
\item Text S1
\item Figures S1 to S8
\item Table S1
\end{enumerate}

\begin{figure}
\centering
\includegraphics[width=0.75\textwidth]{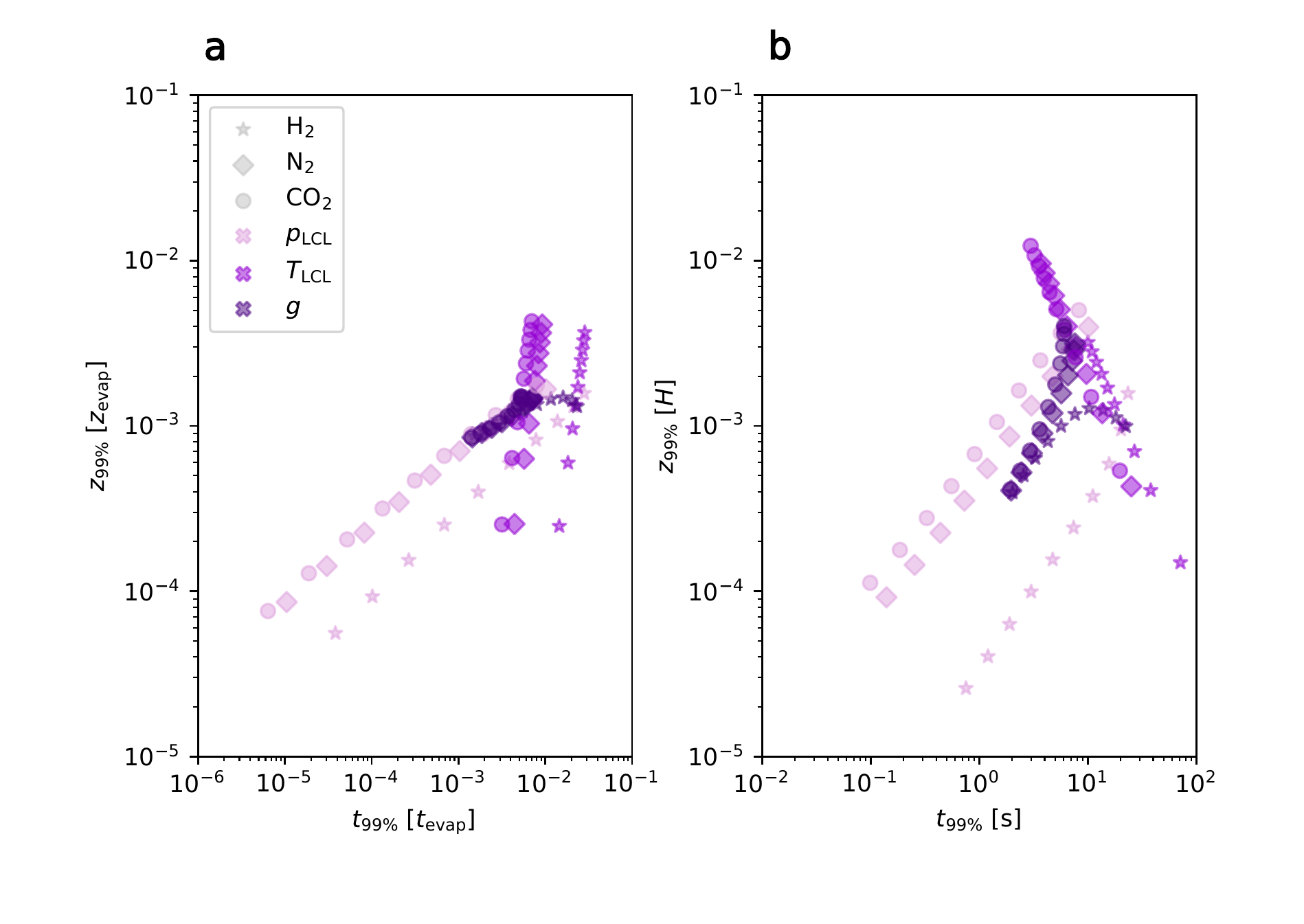}
\caption{(a) Vertical distance from cloud base for raindrop to reach 99\% terminal velocity ($z_{99\%}$) in units of fall distance to total evaporation ($z_\text{evap}$) versus time for raindrop to reach 99\% terminal velocity ($t_{99\%}$) in units of fall time to total evaporation ($t_\text{evap}$) across a broad planetary parameter space. (b) Same as (a) except $z_{99\%}$ given in units of atmospheric scale heights and $t_{99\%}$ in units of seconds. For \ce{H2}, \ce{N2}, and \ce{CO2} background composition atmospheres (marker shape as labeled), we vary in turn $T_\text{LCL}$, $p_\text{LCL}$, and $g$ (color as labeled) over ranges given in Table 1 under ``broad'' from baseline conditions of {$T_\text{LCL}=280$ K}, {$p_\text{LCL}=10^4$ Pa}, and Earth surface gravity. We test 90 different atmospheric conditions with \ce{H2O} raindrops in all cases. Raindrop acceleration is integrated from drag and gravitational forces following Newton's second law. For maximum time to reach terminal velocity, we perform this calculation using {$r_0=r_\text{max}$} and un-physically begin integration at cloud base from {d$z$/d$t$=0 m s$^{-1}$}.}
\label{sfig:95_terminal}
\end{figure}

\begin{figure}
\centering
\includegraphics[width=0.75\textwidth]{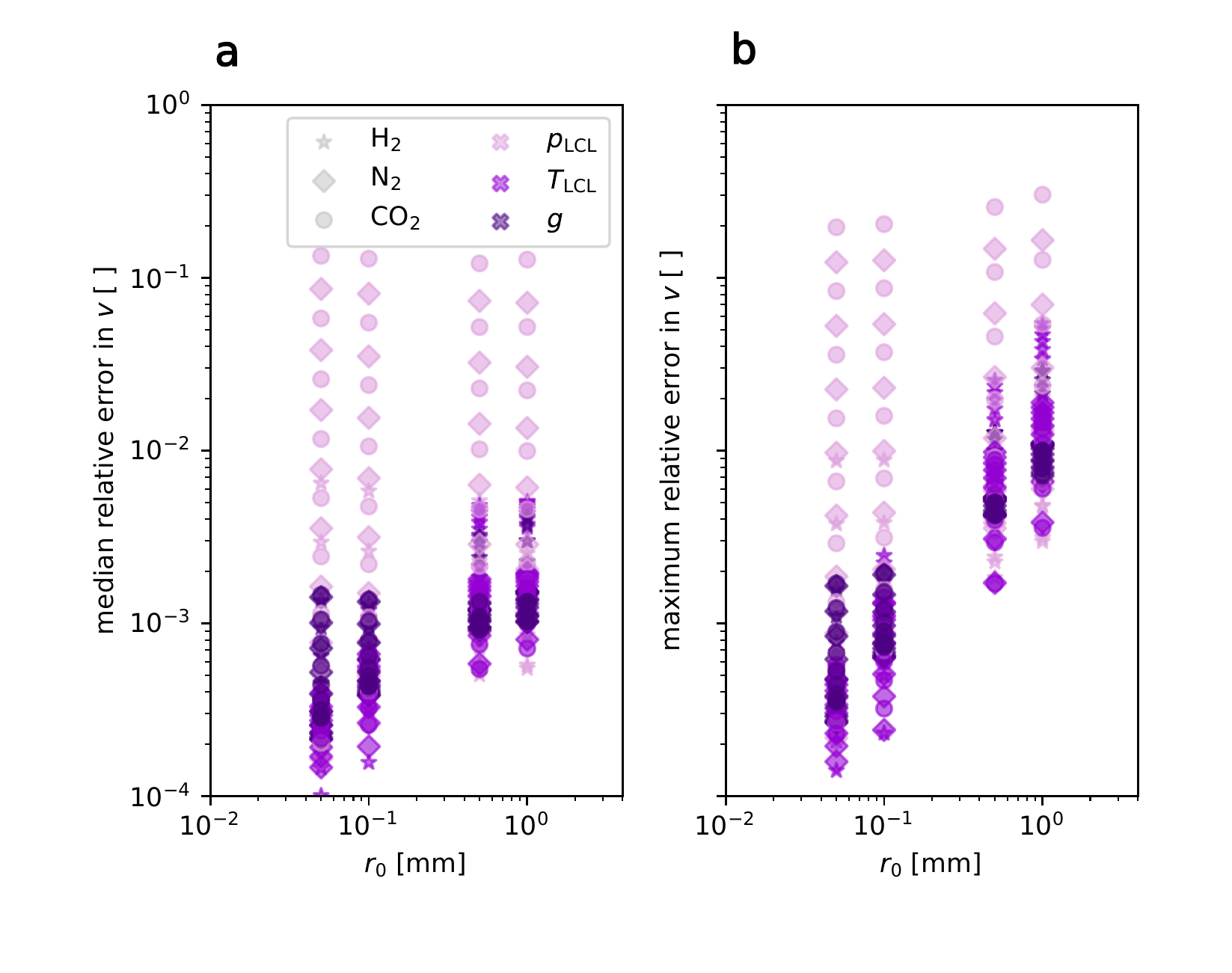}
\caption{(a) Median relative error in raindrop velocity $v$ from assuming terminal velocity is instantly reached during raindrop evaporation (relative to self-consistent treatment of raindrop acceleration) versus equivalent radius at cloud base $r_0$ across a broad planetary parameter space. (b) Same as (a) except for maximum relative error in $v$. For \ce{H2}, \ce{N2}, and \ce{CO2} background composition atmospheres (marker shape as labeled), we vary in turn $T_\text{LCL}$, $p_\text{LCL}$, and $g$ (color as labeled) over ranges given in Table 1 under ``broad'' from baseline conditions of {$T_\text{LCL}=280$ K}, {$p_\text{LCL}=10^4$ Pa}, and Earth surface gravity. We test 90 different atmospheric conditions with \ce{H2O} raindrops in all cases. Raindrop acceleration is integrated from drag and gravitational forces following Newton's second law.}
\label{sfig:terminal_w_evap}
\end{figure}

\begin{figure}
\centering
\includegraphics[width=0.75\textwidth]{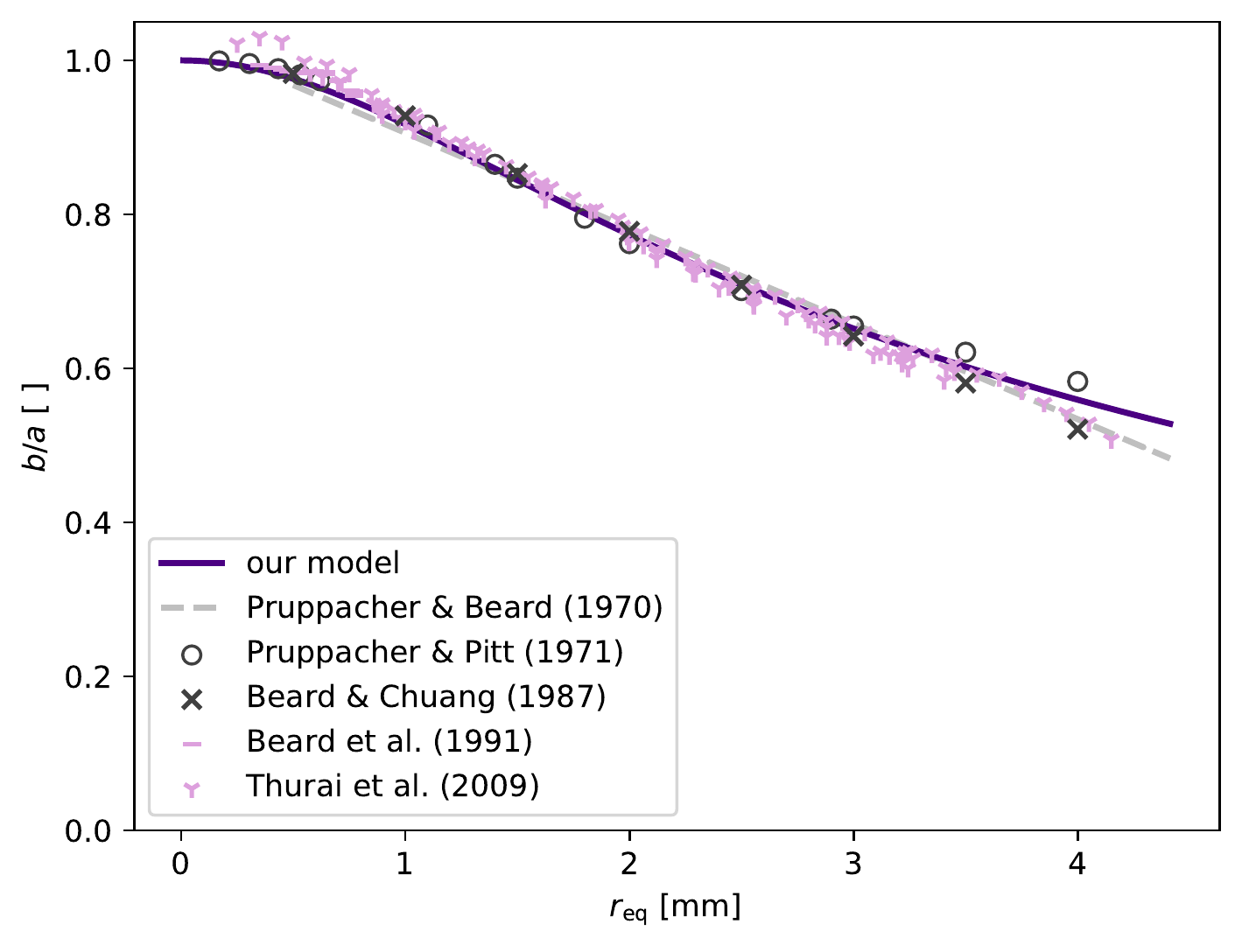}
\caption{Raindrop axis ratio $b/a$ versus equivalent radius $r_\text{eq}$ for \ce{H2O} raindrops and Earth surface gravity. The dark purple line shows our theoretical calculation from equation (2) \cite<following>[]{Green1975}. Dark gray points show results from more comprehensive theoretical models \cite{Pruppacher1971,Beard1987}. Light purple points show experimental results \cite{Beard1991a,Thurai2009}. The dashed light-gray line shows an empirical fit to additional experimental results \cite{Pruppacher1970}.}
\label{sfig:earth_shape}
\end{figure}

\begin{figure}
\centering
\includegraphics[width=0.75\textwidth]{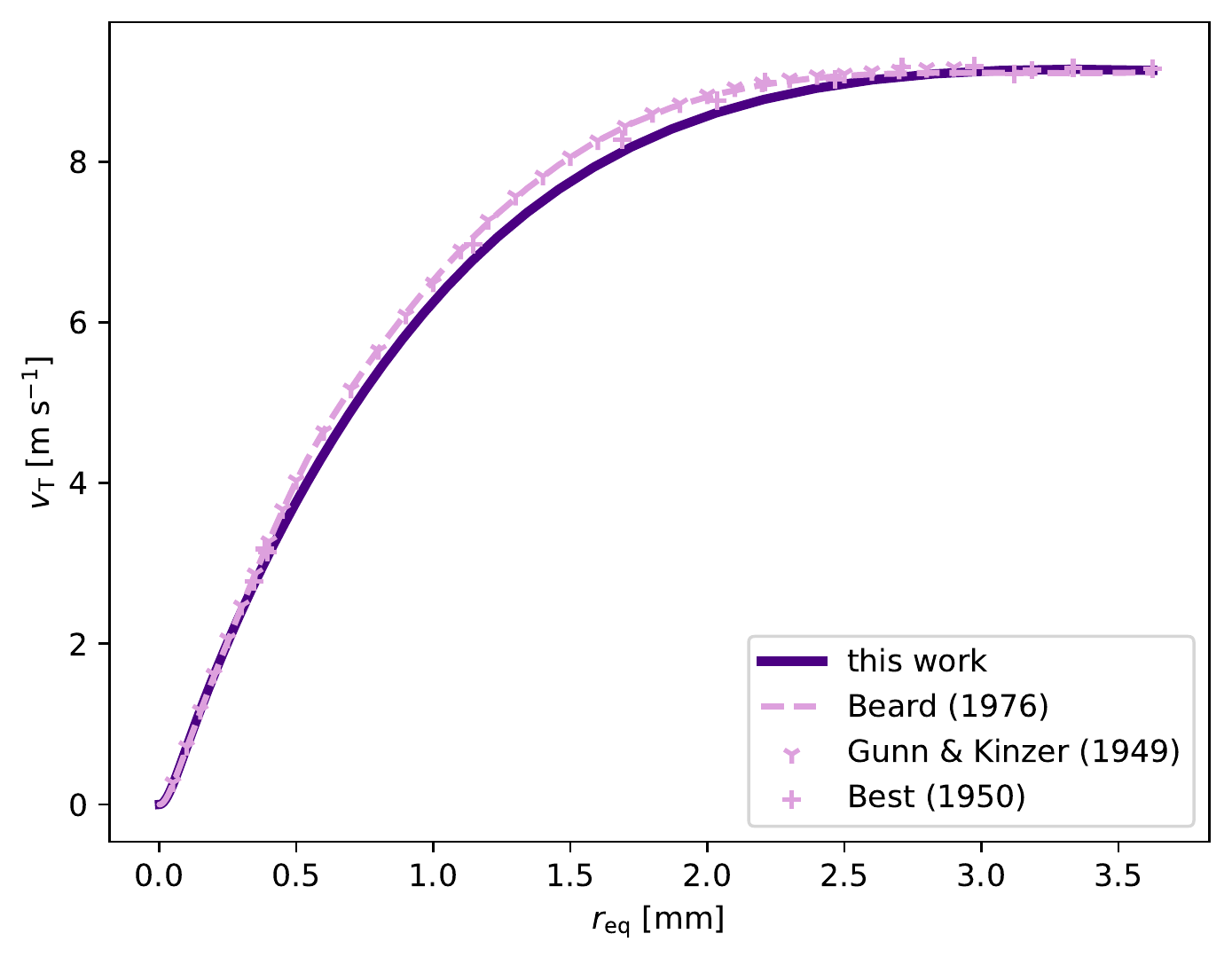}
\caption{Raindrop terminal velocity $v_\text{T}$ versus equivalent radius $r_\text{eq}$ for \ce{H2O} raindrops. Atmospheric parameters are set to mimic Earth experimental conditions: $T=293.15$ K, $p=1.01325\times10^5$ Pa, RH=0.5, Earth surface gravity, and dry air composition of 20\% \ce{O2} and 80\% \ce{N2}. The dark purple line shows our theoretical calculation from equation (8). The light-purple scatter points represent the experimental data of \citeA{Gunn1949} and \citeA{Best1950}. The dashed light-purple line shows our evaluation of the empirical relationship of \citeA{Beard1976}.}
\label{sfig:earth_v}
\end{figure}

\begin{figure}
\centering
\includegraphics[width=0.75\textwidth]{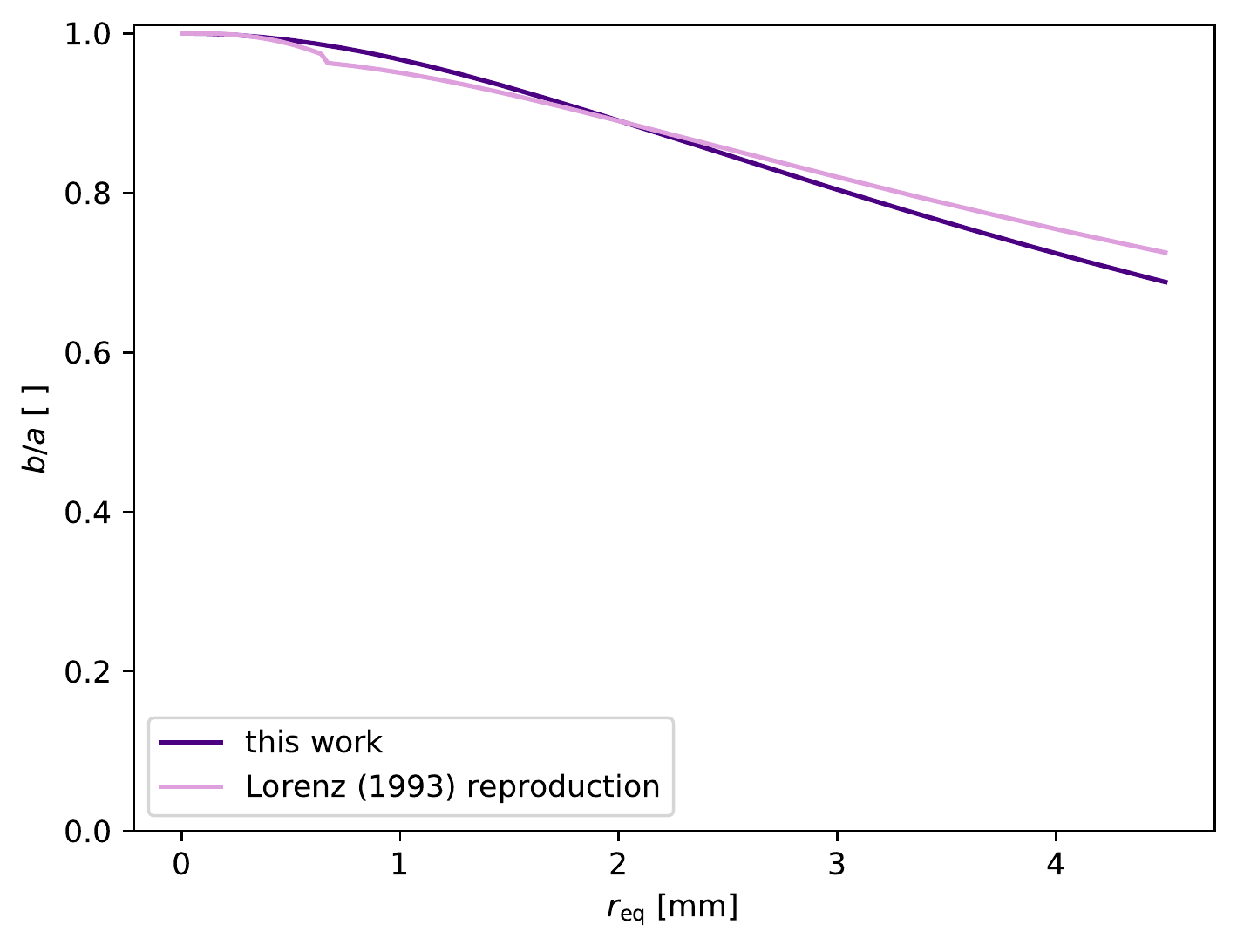}
\caption{Raindrop axis ratio $b/a$ versus equivalent radius $r_\text{eq}$ for \ce{CH4}-\ce{N2} raindrops and Titan surface gravity. The dark-purple line shows our theoretical calculation from equation (2) \cite<following>[]{Green1975}. The light-purple line shows our reproduction of the theoretical method of \citeA{Lorenz1993}. (We do not give scatter points of values directly from \citeA{Lorenz1993} as $b/a$ is calculated from a simple two-component piecewise function.)}
\label{sfig:titan_shape}
\end{figure}

\begin{figure}
\centering
\includegraphics[width=0.75\textwidth]{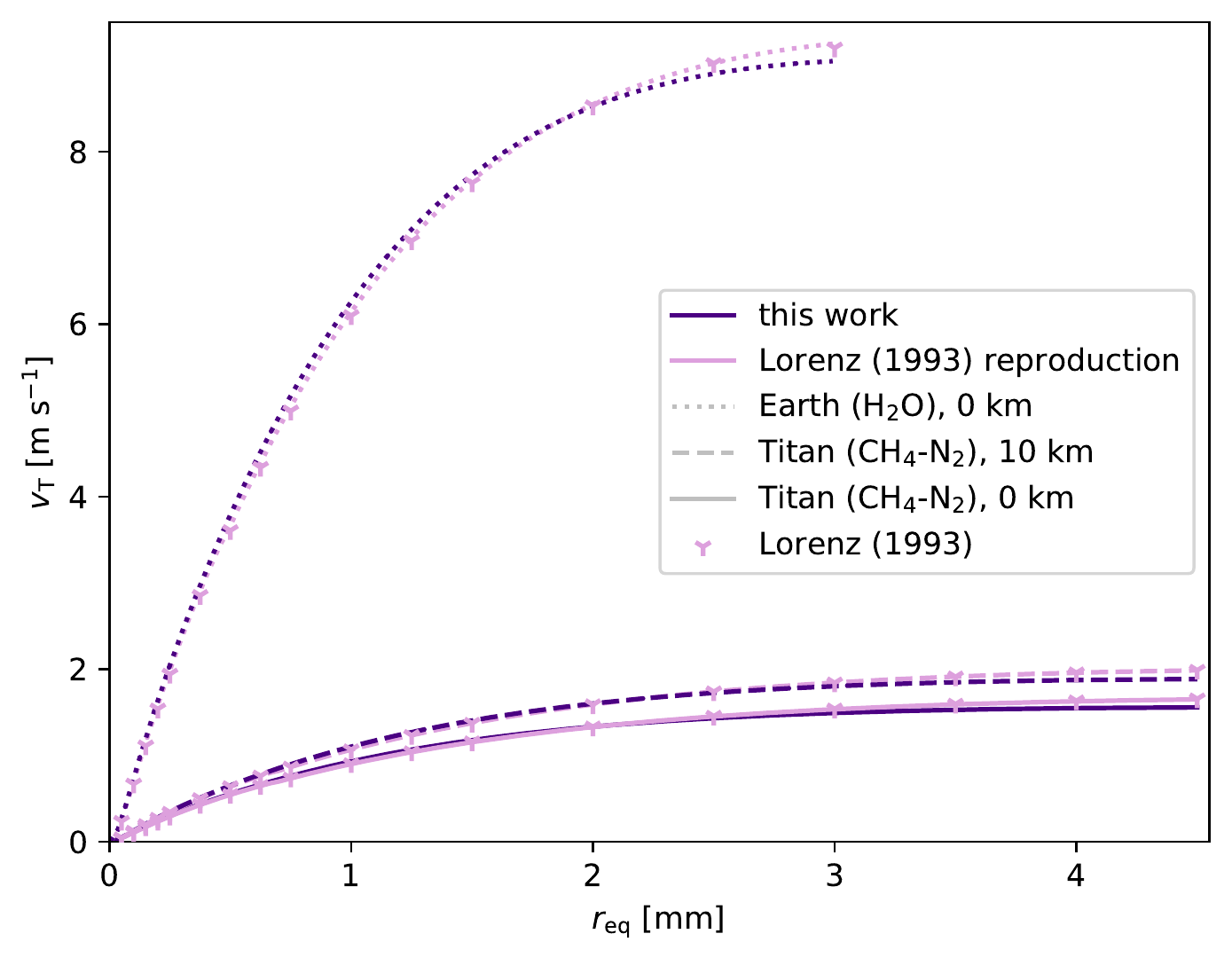}
\caption{Raindrop terminal velocity $v_\text{T}$ versus equivalent radius $r_\text{eq}$ for Earth and Titan conditions (line styles as labeled). The dark purple line shows our theoretical calculation from equation (8). The light-purple line shows our reproduction of the theoretical method of \citeA{Lorenz1993}. The values of light-purple scatter points are taken from Table 1 of \citeA{Lorenz1993}.}
\label{sfig:titan_v}
\end{figure}

\begin{figure}
\centering
\includegraphics[width=0.75\textwidth]{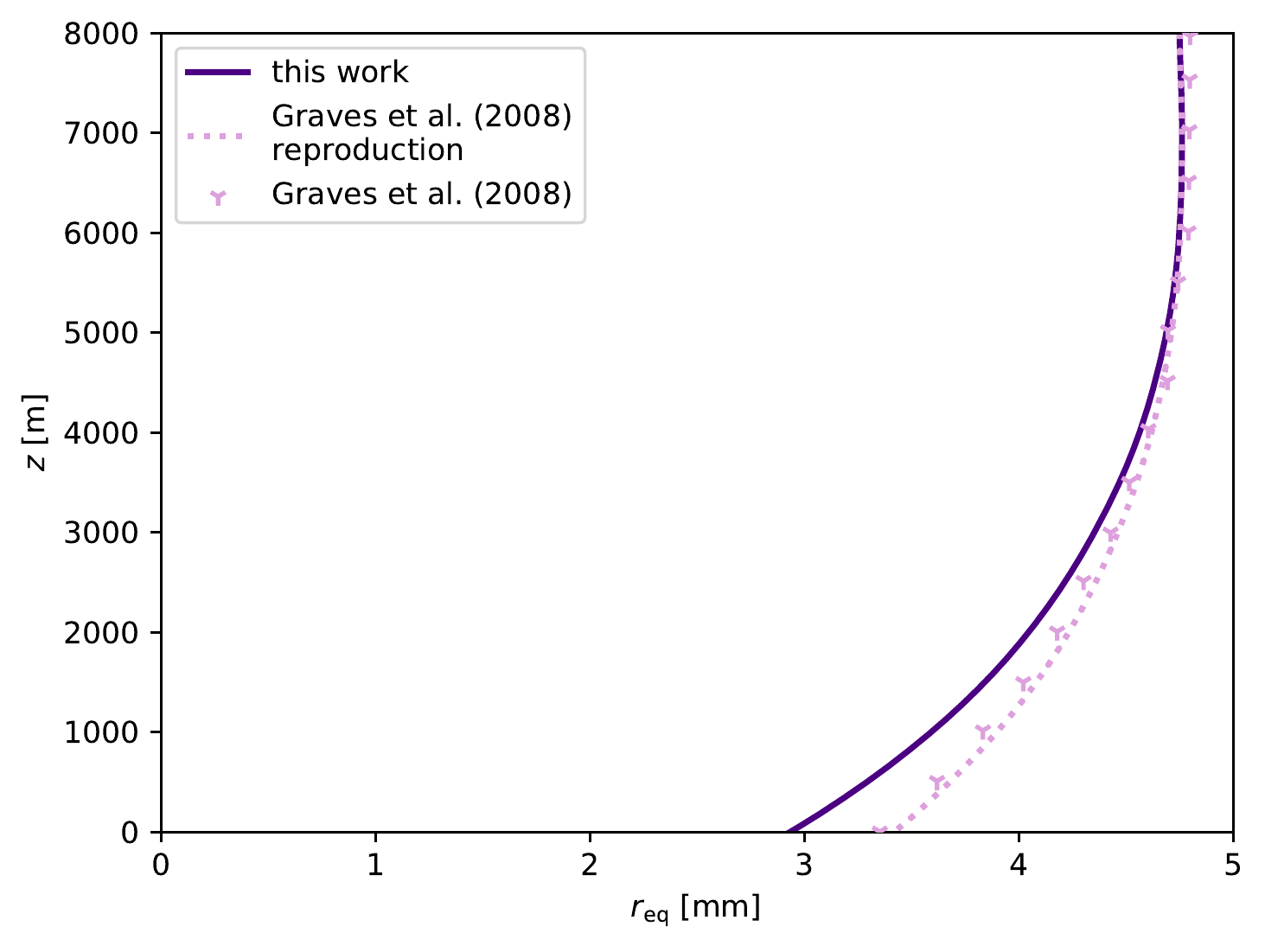}
\caption{Altitude $z$ versus equivalent radius $r_\text{eq}$ for Titan conditions for a \ce{CH4}-\ce{N2} raindrop of initial radius $r_0$=4.75 mm at 8 km. The dark-purple line is calculated following the methods outlined in sections 2-3. The light-purple line is calculated via our reproduction of the methods of \citeA{Graves2008}. The values of light-purple scatter points are taken from Figure 2(b) of \citeA{Graves2008}. From sensitivity tests, the majority of the difference in the calculated $r(z)$ between our method and that of \citeA{Graves2008} is due to our choice to not assume the ventilation factor for heat transport is equal to the ventilation factor for molecular transport. Thus the difference is not a fundamental disagreement of models but rather the result of uncertain parameters quantifiable by fluid dynamics experiments.}
\label{sfig:titan_rz}
\end{figure}

\begin{figure}
\centering
\includegraphics[width=0.75\textwidth]{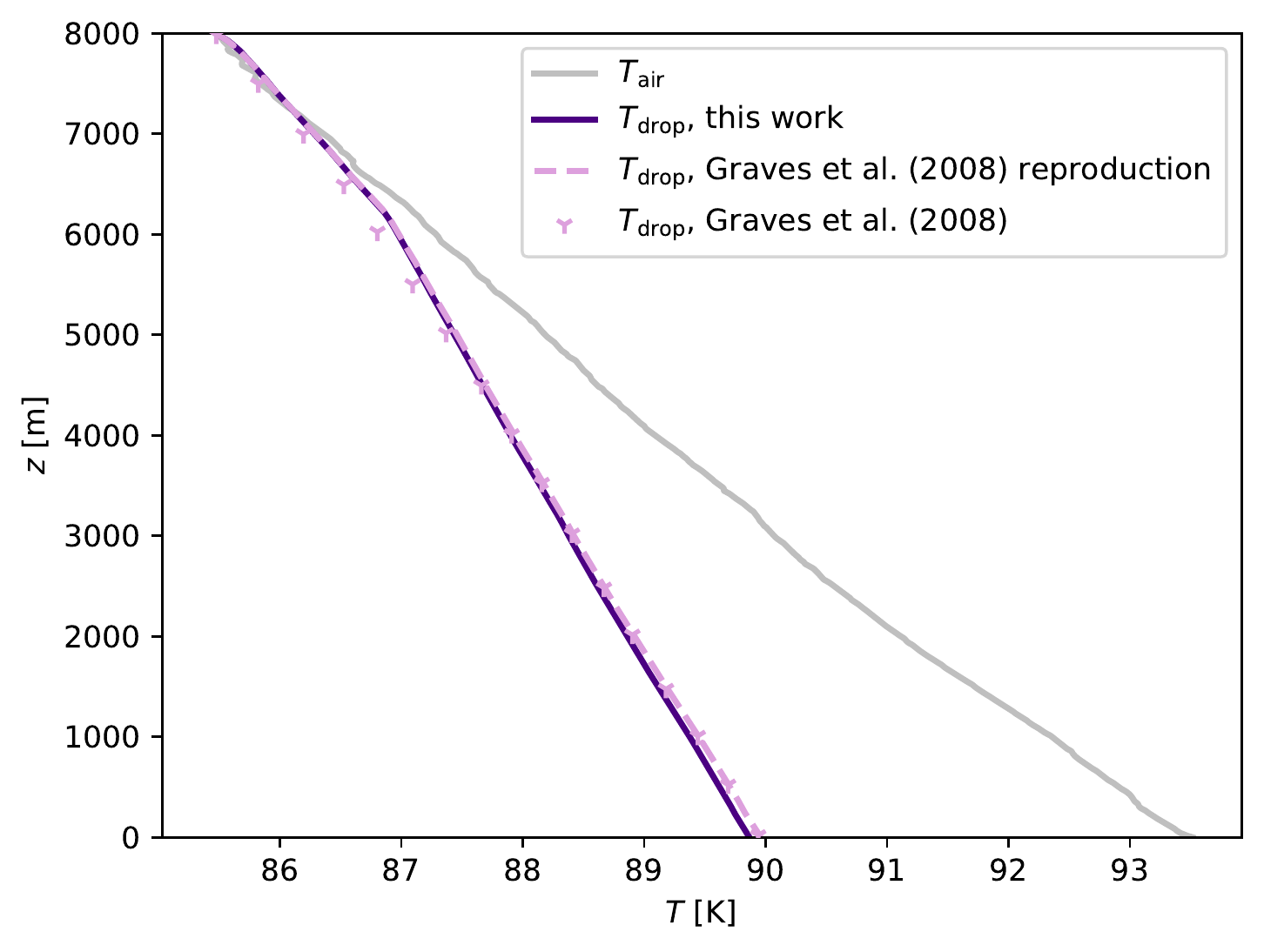}
\caption{Altitude $z$ versus raindrop temperature $T_\text{drop}$ (purple lines) and air temperature $T_\text{air}$ (gray line) for Titan conditions for a \ce{CH4}-\ce{N2} raindrop of initial radius $r_0$=4.75 mm at 8 km. The dark-purple line is calculated following the methods outlined in sections 2-3. The dashed light-purple line is calculated via our reproduction of the methods of \citeA{Graves2008}. The values of light-purple scatter points are taken from Figure 4 of \citeA{Graves2008}. The gray line shows $T_\text{air}$ as recorded by the Huygens probe \cite{Fulchignoni2005}.}
\label{sfig:titan_Tz}
\end{figure}

\noindent\textbf{Introduction}
Figures \ref{sfig:95_terminal}-\ref{sfig:terminal_w_evap} support that raindrop velocity relative to air can be well approximated by raindrop terminal velocity across broad planetary conditions. Figures \ref{sfig:earth_shape}-\ref{sfig:earth_v} show validation tests for our model against observations and empirical relationships for Earth raindrops. Figures \ref{sfig:titan_shape}-\ref{sfig:titan_Tz} and Table \ref{stab:g08_val} compare our model to previous theoretical work focused on Titan raindrops with further description given in Text S\ref{ssec:titan_comp}.

% \clearpage

\noindent\textbf{Text S1.}\label{ssec:titan_comp}
To compare our methods to previous theoretical work on Titan raindrops by \citeA{Lorenz1993} and \citeA{Graves2008}, we present results (1) as reported by the previous authors, (2) from equivalent calculations done by attempting to reproduce their methods within the basic framework of our model (labeled as ``reproduction''), and (3) from equivalent calculations done using the methodology we describe in this work. We note that the scatter points labeled \citeA{Graves2008} in Figures \ref{sfig:titan_rz} and \ref{sfig:titan_Tz} were digitized by us from their figures and are not directly reported values. (We know there is at least one inconsistency in what is plotted in Figures \ref{sfig:titan_rz} as our digitized starting radius is $r_0=4.80$ mm rather than the reported $r_0=4.75$ mm.)

Where possible, we follow condensible and atmosphere properties as described or referenced in each text for the most direct comparisons; we review the most notable ones here. 
Under Titan's atmospheric conditions, \ce{N2} gas is expected to dissolve into liquid \ce{CH4} at significant molar concentrations \cite<e.g.,>[]{Thompson1992}. \citeA{Lorenz1993} accounts for this mixture by prescribing a liquid raindrop density between that expected for pure \ce{CH4} and pure \ce{N2}. \citeA{Graves2008} accounts for this mixture more rigorously: accounting for the \ce{CH4}-\ce{N2} mixture thermodynamics outlined by \citeA{Thompson1992} and evaporation of both \ce{CH4} and \ce{N2} using Raoult's law. 
\citeA{Graves2008} uses pressure, altitude, temperature, and composition data from Cassini's Huygens probe \cite{Fulchignoni2005,Niemann2005}. We linearly interpolate this data to be in terms of altitude to use in our numerical integration. 
% \clearpage
%% ------------------------------------------------------------------------ %%
%%  REFERENCE LIST AND TEXT CITATIONS

%%%%%%%%%%%%%%%%%%%%%%%%%%%%%%%%%%%%%%%%%%%%%%%
% 
%
\bibliography{bib} %do not specify file extension

%% ------------------------------------------------------------------------ %%
%
%  END ARTICLE
%
%% ------------------------------------------------------------------------ %%
\end{article}
% \clearpage

\begin{table}
\settablenum{S1} %%Change number for each table
\caption{Comparison to Table 1 of \citeA{Graves2008}\tablenotemark{a}}
\label{stab:g08_val}
\centering
\begin{tabular}{l c c c}
\hline
& Graves et al. (2008) & Graves et al. (2008)  & this work \\
&  & reproduction & \\
                                   \hline
% $r_0$ [mm]                         & 4.75                 & 4.75                    & 4.75             \\
$r(z=0)$ [mm]                      & 3.34                 & 3.42                    &      2.93            \\
$t_\text{fall}$ [min]              & 78                   & 77                      &            80      \\
$v(z=0)$  [m s$^{-1}$]                         & 1.5                  & 1.5                     &            1.5      \\
$f_\text{\ce{CH4}}(z=0)$ [mol mol$^{-1}$] & 0.77                 & 0.77                    &           0.77       \\
$f_\text{\ce{N2}}(z=0)$ [mol mol$^{-1}$] & 0.23                 & 0.23                    &         0.23         \\
$T_\text{drop}(z=0)$ [K]                & 90.0                 & 90.0                    &            89.9 \\     
\hline
\end{tabular}
\tablenotetext{a}{Properties of a \ce{CH4}-\ce{N2} raindrop of initial size {$r_0=4.75$} mm at 8 km after falling to Titan's surface ($z=0$ km). $t_\text{fall}$ is the total fall time, and $f_\text{X}$ is the liquid molar concentration of species X in the raindrop. In the first column, we give the values reported by \citeA{Graves2008}. In the second column, we show the results our own attempt to reproduce the methods of \citeA{Graves2008}. In the third column, we show the results of our methods for the same calculation. Figures \ref{sfig:titan_rz} and \ref{sfig:titan_Tz} show altitude-dependent properties of the same raindrop.}
\end{table}